\documentclass[12pt,a4paper]{article}
\pdfoutput=1
\usepackage[a4paper]{geometry}

\usepackage{amsmath}
\usepackage{amsfonts}
\usepackage{amssymb}
\usepackage{dsfont}
\usepackage[nosort]{cite}
\usepackage{hyperref}
\usepackage{multirow}

\newcommand{\op}{\hspace{1pt}}

\newcommand{\dd}{\mathrm{d}}

\newcommand{\Op}{\mathcal{O}}

\newcommand{\zb}{\bar{z}}
\newcommand{\Lie}[2]{\mathcal{L}_{#1}#2}

\allowdisplaybreaks[2]
\numberwithin{equation}{section}

\begin{document}
	
	\vspace*{-1.5cm}
	\begin{flushright}
		{\small
			LMU-ASC 42/20
		}
	\end{flushright}
	
	\vspace{1.75cm}
	
	\begin{center}
		{\LARGE
			Asymptotic symmetries in\\ 
			spatially flat FRW\\
		}
	\end{center}
	
	\vspace{0.4cm}
	
	\begin{center}
		Mart\'in Enr\'iquez Rojo, Till Heckelbacher
	\end{center}
	
	\vspace{0.3cm}
	
	\begin{center} 
		\textit{Arnold Sommerfeld Center for Theoretical Physics\\[1pt]
			Ludwig-Maximilians-Universit\"at \\[1pt]
			Theresienstra\ss e 37 \\[1pt]
			80333 M\"unchen, Germany}
	\end{center} 
	
	\vspace{1.8cm}
	
	
	\begin{abstract}
		\noindent
		We perform an off-shell treatment of asymptotically decelerating spatially flat FRW spacetimes at future null infinity. We obtain supertranslation and superrotation-like asymptotic diffeomorphisms which are consistent with the global symmetries of FRW and we compute how the asymptotic data is transformed under them. Further, we study in detail the effect of these diffeomorphisms on some simple backgrounds including unperturbed FRW and Sultana-Dyer black hole. In particular, we investigate how these transformations act on several cosmologically perturbed backgrounds.
	\end{abstract}
	
	
	\clearpage
	
	\tableofcontents
	
	
	\section{Introduction}

	Since its discovery in 1915, General Relativity \cite{Einstein:1916vd} has been extensively explored. Nevertheless, the asymptotic structure of the theory was not investigated until the seminal work of Bondi, van der Burg, Metzner and Sachs (BMS) \cite{Bondi:1962px,Sachs:1962wk}. Contrary to the intuitive idea that one should only recover the Poincar\'e group at future null infinity of asymptotically flat spacetimes, they unveiled a much richer set of asymptotic transformations which also included the so-called supertranslations. The literature around this topic was surrounded by mathematical formality in those days (see e.g. \cite{1960AnPhy..10..171P,PhysRevLett.10.66,Newman:1966ub,Petrov:2000bs,LBell,Sachs:1961zz,Sachs:1962zza,Geroch:1977jn,Wald:1984rg}).

	A couple of decades after BMS, Brown and Henneaux \cite{Brown:1986nw} applied a similar approach to $\text{AdS}_3$, noticing that the algebra of asymptotic diffeomorphisms (and their charges) corresponded to a two dimensional CFT. Their paper was followed by successful attempts to roughly estimate a microscopic description for the BTZ black hole entropy \cite{Strominger:1997eq,Carlip:2000nv,Birmingham:2001dt}, and it was intimately related to the holographic current \cite{tHooft:1993dmi,Susskind:1994vu,Bousso:2002ju} falling into Maldacena's AdS/CFT correspondence \cite{Maldacena:1997re}.

	The modern era of asymptotic symmetries started with the Kerr/CFT correspondence \cite{Guica:2008mu} and the inclusion of superrotations \cite{deBoer:2003vf,Barnich:2009se,Barnich:2010eb}. They were merged with memory effects and soft theorems into infrared triangles \cite{Strominger:2014pwa,Strominger:2017zoo}. It diversified into a wide variety of topics, among which we would like to highlight flat holography \cite{Strominger:2013jfa,Kapec:2014opa,Pasterski:2017ylz,Ball:2019atb,Donnay:2020guq,Donnay:2020fof,Stieberger:2018onx,Fotopoulos:2019vac,Fan:2020xjj}, black hole entropy \cite{Hawking:2016msc,Haco:2018ske,Grumiller:2019fmp,Grumiller:2019ygj,Averin:2016ybl,Averin:2016hhm,Lust:2017gez,Donnay:2015abr,Donnay:2019jiz,Penna:2018bzj}, algebraic oriented studies   \cite{Banerjee:2015kcx,Campoleoni:2015qrh,Campoleoni:2016vsh,Safari:2019zmc,Borowiec:2020ddg}, extension to timelike \cite{H:2020eei} and spatial flat infinity \cite{Prabhu:2019daz}, to $\text{dS}_4$ and $\text{AdS}_4$ \cite{Compere:2020lrt}, to string theory \cite{Bianchi:2014gla,Avery:2015gxa,Afshar:2018sbq,Esmaeili:2020eua} and the swampland \cite{Bonnefoy:2019nzv}, to higher dimensions \cite{Kapec:2015vwa,Hollands:2016oma,Colferai:2020rte} and Kaluza-Klein \cite{Lu:2019jus}, to the membrane paradigm \cite{Penna:2015gza,Penna:2017bdn,Carrillo-Gonzalez:2018wrh}, to alternative gravity theories \cite{Gonzalez:2012nv,Zhang:2013mva,Lu:2020mjp} and also to cosmological settings. 
	
	Surprisingly, the literature regarding the infrared structure of cosmology is scarce \cite{Shiromizu:1999iq,Hinterbichler:2012nm,Hinterbichler:2013dpa,Ferreira:2016hee,Mirbabayi:2016xvc,Tolish:2016ggo,Hamada:2017gdg,Donnay:2019zif}. Our work is framed in this context with delving into the not yet well-understood \textit{asymptotic symmetry corner of the cosmological infrared triangle} being our main objective. More concretely, we study asymptotically decelerating spatially flat Friedmann-Robertson-Walker (FRW) spacetimes at future null infinity $\mathcal{I}^{+}$ using an off-shell formalism. Our approach is therefore not restricted to General Relativity and can be applied to alternative gravity theories which include FRW as a solution. We define the class of metrics to be considered asymptotically decelerating spatially flat FRW, without neither explicitly fixing matter content nor establishing relationships among the metric coefficients in the large $r$-expansion. An on-shell analysis is the natural step to pursue in future studies.
	
	Our universe is not asymptotically flat, therefore, we aim to translate a BMS-like analysis into a more phenomenological framework. Even though experimental data suggests that we live in a FRW with accelerating expansion \cite{Riess:1998cb}, we chose a decelerating FRW to have a boundary at null infinity. As our universe went through a phase of decelerated expansion, we can imagine ourselves as observers looking at a decelerated universe from null infinity. This is only true as an approximation and we would have to extend our analysis to accelerated FRW spacetimes to get exact realistic results. 
	
	The price of going from asymptotically flat to asymptotically FRW is that now we have to consider a time-dependent boundary metric instead of the simpler time-independent Minkowski. However, we manage to obtain consistent supertranslation and superrotation-like tranformations and their action on the asymptotic data. We consistently recover the asymptotically flat results in the appropriate limit. Furthermore, we apply these diffeomorphisms to unperturbed FRW, Sultana-Dyer and cosmologically perturbed FRW backgrounds. These examples already permit us to appreciate the advantages, as well as to discover the limitations of our approach.

	Let us now shortly discuss the unique works to date which tackle this class of spacetimes: 
	
	\begin{itemize}
		\item The first work delving into the infrared structure of decelerating spatially flat FRW at null infinity is \cite{Kehagias:2016zry}. Concretely, they focused on the cosmological gravitational memory effect and analyzed BMS transformations on FRW spacetimes. While recognizing their pioneer labor, the authors made unfortunate computational mistakes which question most of the analysis at the quantitative level. More fundamentally, they considered flat BMS transformations as asymptotic symmetries of FRW spacetimes, which they are not because they do not respect their isometries. Besides, they incorrectly denominated several coefficients as cosmological modes when they do not transform as such under rotations. Whereas supertranslations were considered, we extend the analysis to superrotations with a more general ansatz.
		\item On the course of the elaboration of this paper, \cite{Bonga:2020fhx} was published. The authors followed a more formal approach than ours, using the covariant formalism of a conformal completion \`a la Penrose. It might be better suited for some features, like an on-shell description, while ours is more straightforward for practical usage and application to examples. They have also noted the inconsistencies in \cite{Kehagias:2016zry}, and they did not study superrotations. It would be interesting to explore the compatibility between both works and how they can be complemented.
	\end{itemize}

	This paper is organized as follows: in section~\ref{sec_aflat} we shortly review asymptotically flat spacetimes. In section~\ref{sec_asfFLRW} we analyze asymptotic diffeomorphisms on perturbed spatially flat FRW spacetimes. This comprises to define the asymptotic metrics we consider, to obtain the supertranslation and superrotation-like transformations which preserve them, and to analyze their effect on the metric coefficients. In section~\ref{sect_examples} we discuss various background examples, and section~\ref{sect_conc} contains our conclusions. In two appendices, we have collected the Einstein equations and the Lie derivatives for the asymptotic metrics.

	
	\section{Review of asymptotically flat spacetimes}
	\label{sec_aflat}
	
	In this section we briefly review asymptotically flat spacetimes at future null infinity $\mathcal{I}^{+}$, paving the way for the analysis in section \ref{sec_asfFLRW}. The following discussion is mostly based on \cite{Strominger:2017zoo,Hamada:2017gdg,Campiglia:2014yka,Compere:2018aar}.


	\subsubsection*{General procedure}
	\label{sec_genproc}

	Since the pioneer works of Bondi, van der Burg, Metzner \cite{Bondi:1962px} and Sachs \cite{Sachs:1962wk}, many studies have been performed allowing for different falloff conditions on the metric and on the diffeomorphisms generating the asymptotic transformations. A common feature most of these approaches share is the use of Bondi coordinates
	\begin{equation}
		u=t-\sqrt{x^ix_i} \ , \ \ r=\sqrt{x^ix_i} \ , \ \ z=\frac{x^1+ix^2}{x^3+\sqrt{x^ix_i}} \ , \ \ \zb=\frac{x^1-ix^2}{x^3+\sqrt{x^ix_i}} \ ,
	\end{equation}
	adequate to describe the asymptotic metrics near $\mathcal{I}^{+}$, together with the Bondi gauge
	\begin{equation}
		g_{rr}=g_{rA}=0 \ , \ \ \partial_r\det\left(\frac{g_{AB}}{r^2}\right)=0 \ ,
		\label{eq:BG}
	\end{equation}
	which completely fixes the local diffeomorphism invariance.
	
	Nevertheless, we still need to specify what we consider by asymptotic flatness. This is accomplished through a choice of falloff conditions on the metric components at large $r$. The asymptotic symmetries are generated by diffeomorphisms that preserve the Bondi gauge (\ref{eq:BG}), as well as the selected boundary conditions. Therefore, a final ingredient is the large-$r$ falloff conditions on the diffeomorphisms.


	\subsection{Supertranslations}
	\label{aflatstran}
	
	Supertranslations are derived from a rather restrictive choice of boundary conditions, which still allows for interesting physical solutions, given by
	\begin{align}
		g_{uu}=-1+\mathcal{O}(r^{-1}) \ ; \ \ g_{ur}=-1+\mathcal{O}(r^{-2}) \ ; \ \ g_{uz}=\mathcal{O}(1)\\
		g_{zz}=\mathcal{O}(r) \ ; \ \ g_{z\zb}=r^2\gamma_{z\zb}+\mathcal{O}(1) \ ; \ \  g_{rr}=g_{rz}=0
		\label{eq:S1}
	\end{align}
	and 
	\begin{equation}
		\xi^u, \xi^r\sim \mathcal{O}(1) \ , \ \ \xi^{z}, \xi^{\zb} \sim \mathcal{O}(r^{-1}) \ .
		\label{eq:falloffstrans}
	\end{equation}
	
	At large $r$, the structure of the metric is constrained to be of the form \footnote{ \label{note1} Assuming the falloff conditions (2.6) in \cite{Flanagan:2015pxa} for the stress energy tensor and using Einstein equations. Such assumptions are motivated by the behaviour of radiative scalar field solutions in Minkowski spacetime.}
	\begin{align}
		\dd s^2=&-\left(1-\frac{2m}{r}\right)\dd u^2-2\dd u\dd r
		+D^zC_{zz}\dd u\dd z+D^{\zb}C_{\zb\zb}\dd u\dd\zb
		+\frac{4r^2}{(1+z\zb)^2}\dd z\dd\zb \nonumber \\
		&+rC_{zz}\dd z^2+rC_{\zb\zb}\dd\zb^2
		+\frac{1}{r}\left( \frac{4}{3}(N_z+u\partial_{z}m)-\frac{1}{4}\partial_z(C_{zz}C^{zz}) \right)\dd u \dd z+\text{c.c.} + ...
		\label{eq:aflatmetric}
	\end{align}
	and the asymptotic killing vectors are given by
	\begin{align}
		\xi(f(z,\zb) )=&f\partial_u
		+D^AD_A f\partial_r-\frac{1}{r}D^Af\partial_A+... \ \ \ \ A=z,\zb
		\label{eq:StransAF}
	\end{align}
	
	Note from (\ref{eq:StransAF}) that the asymptotic killing vectors are fully determined by smooth functions on the sphere $f(z,\zb)\in\mathcal{C}^{\infty}(S^2)$. Expanding $f$ in spherical harmonics, one can prove that the modes $l=0,1$ correspond (to leading order) respectively to time and space translation generators \cite{Strominger:2017zoo}, while the remaining modes can be roughly interpreted as "angle-dependent translations associated to the conservation of energy at every angle" \cite{Strominger:2017zoo,Penna:2015gza}.


	\subsection{Superrotations}
	\label{aflatsrot}
	
	The transformations can be enlarged relaxing some of the previous conditions, being still the solutions physically acceptable. We consider now diffeomorphisms with the following large $r$ behaviour 
	\begin{equation}
		\xi^u, \xi^r\sim \mathcal{O}(1) \ , \ \ \xi^{z}, \xi^{\zb} \sim \mathcal{O}(1) \ .
		\label{eq:falloffsrot}
	\end{equation}
	
	The novel $\mathcal{O} (1)$ modes in $\xi^A$ allow for extra asymptotic symmetries naturally associated to rotations and boosts:
	\begin{align}
		\xi(V^A(z,\zb))=&V^A\partial_A
		+\frac{u}{2}D_AV^A\partial_u-\frac{r}{2}D_AV^A\partial_r \nonumber \\
		&-\frac{u}{2r}D^AD_BV^B\partial_A+\frac{u}{4}D_BD^BD_AV^A\partial_r+ ...
		\label{eq:SrotAF}
	\end{align}
	
	Therefore, the enhanced asymptotic symmetries are given by:
	\begin{equation}
		\xi(f,V^A)=\xi(f)+\xi(V^A) \ .
		\label{eq:fullAF}
	\end{equation}
	
	Nevertheless, the previous falloff conditions for the metric (\ref{eq:S1}) are not generally preserved under (\ref{eq:SrotAF}). Concretely, terms with $\delta g_{AB}\sim \mathcal{O}(r^2)$ and $\delta g_{uu}\sim \mathcal{O}(1)$ arise. This leads us to the following possibilities analyzed in the literature:
	
	\begin{itemize}
		\item \textit{$SO(1,3)\ltimes\mathcal{C}^{\infty}(S^2)$} \cite{Bondi:1962px,Sachs:1962wk}
		
		If the only allowed superrotations are $V^A=1,z,z^2,i,iz,iz^2$, that is, the six global CKV on $S^2$, one can show to leading order in $r$ that (\ref{eq:SrotAF}) generates the Lorentz transformations \cite{Strominger:2017zoo}. Therefore, together with the $l=0,1$ modes of $f$ in (\ref{eq:StransAF}), we recover the Poincaré algebra from (\ref{eq:fullAF}). Terms with $\delta g_{AB}\sim \mathcal{O}(r^2)$ and $\delta g_{uu}\sim \mathcal{O}(1)$ do not show up.

		\item \textit{$\text{Conf}(S^2)\ltimes\mathcal{C}^{\infty}(S^2)$} \cite{deBoer:2003vf,Barnich:2009se,Barnich:2010eb,Strominger:2016wns}
		
		Another natural possibility is to admit also the locally defined CKV on $S^2$. Terms with $\delta g_{AB}\sim \mathcal{O}(r^2)$ arise only at isolated points corresponding to the singularities of the meromorphic CKV in a similar fashion as found long ago in 2d CFT \cite{Belavin:1984vu}.
		
		\item \textit{$\text{Diff}(S^2)\ltimes\mathcal{C}^{\infty}(S^2)$} \cite{Campiglia:2014yka,Campiglia:2015yka}
		
		A broader possibility is to consider all the diffeomorphisms on $S^2$ \cite{Campiglia:2014yka}. Terms with $\delta g_{AB}\sim \mathcal{O}(r^2)$ and $\delta g_{uu}\sim \mathcal{O}(1)$ are present. Expanding $V^A\in\text{Vect}(S^2)$ in vector spherical harmonics, one can prove that the modes $l=1$ correspond (to leading order) respectively to the six global CKV generators on $S^2$ \cite{Strominger:2017zoo}, while the remaining modes can be roughly interpreted as "angle-dependent rotations associated to the conservation of momentum at every angle" \cite{Strominger:2017zoo,Penna:2015gza}. 
	\end{itemize}
	
	Let us remark that the Poincaré generators, which are the GKV of Minkowski, arise naturally as the global modes of the asymptotic diffeomorphisms in asymptotically flat spacetimes for the three cases.

	
	\section{Asymptotically spatially flat FRW spacetimes}
	\label{sec_asfFLRW}
	
	We now turn to analyze asymptotically spatially flat FRW spacetimes at $\mathcal{I}^{+}$. 
	
	
	\subsection{Asymptotic metric}
	\label{asflatFLRWam}
	
	Spatially flat FRW spacetimes are conformal to flat spacetimes. In Bondi coordinates 
	\begin{equation}
		u=\eta-\sqrt{x^ix_i} \ , \ \ r=\sqrt{x^ix_i} \ , \ \ z=\frac{x^1+ix^2}{x^3+\sqrt{x^ix_i}} \ , \ \ \zb=\frac{x^1-ix^2}{x^3+\sqrt{x^ix_i}} \ ,
	\end{equation}
	their metrics can be written as
	\begin{equation}
		\dd s^2=\left(\frac{r+u}{L}\right)^{2k}\left(-\dd u^2-2\dd u\dd r+\frac{4r^2}{(1+z\zb)^2}\dd z\dd\zb\right) \ ,
	\end{equation}
	where $\eta$ is the conformal time and $k=2/(3\omega+1)$, being $\omega=p/\rho$ the equation of state parameter of the fluid. 
	
	These spacetimes can be divided into accelerated ($-1<\omega<-1/3$, $k<0$) and decelerated ($-1/3<\omega<1$, $k>0$) expansion. It turns out that only the decelerated have future null infinity $\mathcal{I}^{+}$, such that our analysis restricts to them. Further discussion on the conformal structure of spatially flat FRW in the context of asymptotic symmetries can be found in \cite{Kehagias:2016zry,Bonga:2020fhx}.
	
	The next step is to define the class of metrics to be considered as asymptotically spatially flat FRW. The approach we follow here is to include the most general possible terms compatible with the Bondi gauge and frame (round metric on $S^2$), which allow for scalar, vector and tensor perturbations \footnote{More details concerning cosmological perturbations will come in section \ref{sect_cosmopert}.} but do not change to leading order the characteristic homogeneous, isotropic and spatially flat FRW profile, as well as its matter content (codified in a constant $k$), when $r\to\infty$. A consistency check is that the transformed metrics, generated by general asymptotic diffeomorphisms, close in the $r$ expansion. Moreover, one can test by means of Einstein equations (appendix \ref{einsteintensor}) that the trace and components of the energy momentum tensor generated by such terms remains at most finite when appropriately scaled. This leads to the asymptotic metrics \footnote{A similar asymptotic metric, but derived in a different manner, can be found in \cite{Kehagias:2016zry}.}: 
	\begin{align}
		\dd s^2=&\left(\frac{r+u}{L}\right)^{2k}
		\left\{-\left(1-\Phi-\frac{2m}{r}\right)\dd u^2-2\left(1-\Psi-\frac{K}{r}\right)\dd u\dd r \right.\nonumber\\
		&\left.-2(r\Theta_z+U_z+\frac{N_z}{r})\dd u\dd z-2(r\Theta_{\zb}+U_{\zb}+\frac{N_{\zb}}{r})\dd u\dd\zb
		\right. \nonumber\\
		&\left.+2\left(\frac{2r^2(1+\Omega)}{(1+z\zb)^2}+h_{z\zb}\right)\dd z\dd\zb+(rC_{zz}+h_{zz})\dd z^2+(rC_{\zb\zb}+h_{\zb\zb})\dd\zb^2\right\} \nonumber \\
		=&a^{2}\left\{-\left(1-\Phi-\frac{2m}{r}\right)\dd u^2-2\left(1-\Psi-\frac{K}{r}\right)\dd u\dd r-2(r\Theta_A+U_A+\right. \nonumber \\
		&\frac{1}{r}N_A)\dd u\dd x^A+\left.\left((1+\Omega)r^2\gamma_{AB}+rC_{AB}+h_{AB}\right)\dd x^A\dd x^B\right\} \ .
		\label{eq:asymptpertspatflatFLRW}
	\end{align}
	
	Ultimately, the metrics (\ref{eq:asymptpertspatflatFLRW}) should verify Einstein equations. In general, this would introduce extra constraints and relations between the a priori independent parameters in the $r$-expansion. Nevertheless, if there are no restrictions on the allowed matter and the energy momentum tensor is unconstrained, Einstein equations do not lead to such relationships. To actually perform an on-shell analysis, we would need to constrain and classify the allowed matter with the corresponding falloff conditions. Although it has been accomplished for asymptotically flat spacetimes (c.f. footnote \ref{note1}), it is more subtle in our setting where we cannot restrict to isolated matter distributions, but instead the whole spacetime is filled with hydrodynamical content.
	
	Furthermore, one should explicitly check that the new metrics obtained after application of the asymptotic diffeomorphisms also verify Einstein equations. Such a task is cumbersome, mainly due to the fact that we are dealing with metrics corresponding in general to infinite expansions in $r$, which would require to check Einstein equations order by order. We are not aware of the existence of the mathematical machinery to perform such an analysis for a complex class of metrics like ours \footnote{Technical tools were recently developed for simpler asymptotic metrics, like $AdS_3$ \cite{Compere:2015knw,Seraj:2016cym,Grumiller:2019ygj} due to its topological nature. Nevertheless, even for asymptotically flat spacetimes in four dimensions, such instruments, to our knowledge, are not available.}. 
	
	Our first goal is to find and understand the more general class of asymptotic diffeomorphisms acting on and relating to such a general class of cosmological metrics (\ref{eq:asymptpertspatflatFLRW}), regardless the matter content. Therefore, we consider this final step beyond the purpose of this paper and leave it for future research.
	
	Before we continue, let us note that a different approach was independently developed in \cite{Bonga:2020fhx}, where the construction is more geometric and closer to the original BMS analysis. It would be interesting to investigate to which extent our results are compatible.

	
	\subsection{Supertranslations}
	\label{asflatFLRWstrans}
	
	Following the conventional analysis in flat space, we begin by studying supertranslations alone
	\begin{align}
		\xi=\xi^u(u,r,z,\zb)\partial_u+\sum\limits_{n=0}^{\infty}\frac{\xi^{r(n)}}{r^n}\partial_r
		+\sum\limits_{n=1}^{\infty}\frac{\xi^{z(n)}}{r^n}\partial_z
		+\sum\limits_{n=1}^{\infty}\frac{\xi^{\zb(n)}}{r^n}\partial_{\zb} \ .
		\label{eq:astrans}
	\end{align}
	
	
	\subsubsection{General case}
	
	The starting point consists of imposing the Bondi gauge. We observe from (\ref{eq:bondigauge1}) and $0=a^{-2}\mathcal{L}_{\xi}g_{rr}$ that $\partial_r\xi^u=0$. Consequently, the general ansatz for supertranslations (\ref{eq:astrans}) becomes
	\begin{align}
		\xi=\xi^u(u,z,\zb)\partial_u+\sum\limits_{n=0}^{\infty}\frac{\xi^{r(n)}}{r^n}\partial_r
		+\sum\limits_{n=1}^{\infty}\frac{\xi^{z(n)}}{r^n}\partial_z
		+\sum\limits_{n=1}^{\infty}\frac{\xi^{\zb(n)}}{r^n}\partial_{\zb} \ .
		\label{eq:astransref}
	\end{align}
	
	The next step is to require $0=a^{-2}\mathcal{L}_{\xi}g_{rA}$, which together with (\ref{eq:liestransrA}) leads to
	\begin{align}
		\xi^{B(1)}=&-\frac{(1-\Psi)}{1+\Omega}D^B\xi^u\label{eq:xi1} \ , \\
		\xi^{B(2)}=&\frac12\left(\frac{1-\Psi}{(1+\Omega)^2}C^{AB}D_A\xi^u+\frac{K}{1+\Omega}D^B\xi^u\right) \ .
		\label{eq:xi2}
	\end{align}
	
	To fix the last gauge condition we have to demand that
	\begin{align}
		\partial_r\det\left(\frac{g_{AB}+\Lie{\xi}{g_{AB}}}{a^2 r^2}\right)=0 \ .
		\label{eq:detBondisutra} 
	\end{align}
	
	We can now expand the determinant around the metric (\ref{eq:asymptpertspatflatFLRW}) with an infinitesimal perturbation $\Lie{\xi}{g_{AB}}$ \eqref{eq:LieAB1}-\eqref{eq:LieAB11}
	\begin{align}
		\det\left(\frac{g_{AB}+\Lie{\xi}{g_{AB}}}{a^2 r^2}\right)=&\det(g)\det\left(\mathbb{I}+\frac{1}{a^2r^2}g^{AC}\Lie{\xi}{g_{CB}}\right)\nonumber \\
		=&\det(g)\left[1+\frac{F^A_A}{1+C}+
		\frac{1}{r}\left(\frac{S^A_A}{1+\Omega}-\frac{C^{AC}F_{CA}}{(1+\Omega)^2}\right)\right.\nonumber \\
		&\left.+\frac{1}{r^2}\frac{1}{1+\Omega}\left(K^A_A-\frac{C^{AB}S_{AB}}{1+\Omega}+\frac{C^A_CC^{CB}F_{AB}}{(1+\Omega)^2}-\frac{h^{AB}F_{AB}}{1+\Omega}\right)\right] 
		\label{eq:explicitbdgauge}
	\end{align}
	such that for \eqref{eq:detBondisutra} to be obeyed we need to require the terms $\mathcal{O}(r^{-1})$ and $\mathcal{O}(r^{-2})$ to vanish. This leads to
	\begin{align}
		\xi^{r(0)}=&\frac{1}{2(1+k)(1+\Omega)}\left[-D_A\left((1+\Omega)\xi^{A(1)}\right)-2k(1+\Omega)\xi^u-\Theta^AD_A\xi^u\right] \ ,
		\label{eq:coeffstransr0} \\
		\xi^{r(1)}=&\frac{1}{2(1+k)(1+\Omega)}\left[2ku(1+\Omega)(\xi^u+\xi^{r(0)})+\frac{C^{AB}\Theta_AD_B\xi^u}{1+\Omega}\right.\nonumber\\
		&\left.-U^AD_A\xi^u-D_A((1+\Omega)\xi^{A(2)})\right] \ .
		\label{eq:coeffstransr1}
	\end{align}
	
	The remaining Lie derivatives (\ref{eq:Lieuu})-(\ref{eq:LieuA}) close in the $r$ expansion (\ref{eq:asymptpertspatflatFLRW}). As a consequence, we do not get more conditions.
	
	\paragraph{Summary of results}
	\begin{enumerate}
		\item Supertranslations have the form:
		\begin{align}
			\xi=&\xi^u(u,z,\zb)\partial_u+\left[\xi^{r(0)}+\frac{1}{r}\xi^{r(1)}\right]\partial_r+\left[\frac1r\xi^{B(1)}+\frac{1}{r^2}\xi^{B(2)}\right]\partial_B \ .
			\label{eq:Stranslac}
		\end{align}
		\item The only free parameter is $\xi^u(u,z,\zb)$.
		\item The other coefficients in \eqref{eq:Stranslac} are given by (\ref{eq:coeffstransr0}), (\ref{eq:coeffstransr1}), (\ref{eq:xi1}) and (\ref{eq:xi2}).
		\item From the Lie derivatives in appendix \ref{liederstrans}, we observe that the action of \eqref{eq:Stranslac} satisfies the falloff conditions (\ref{eq:asymptpertspatflatFLRW}) automatically, such that we do not get any additional restrictions on the supertranslations, and induces the following transformations in the asymptotic data:
		\begin{align}
			\delta\Phi=&\xi^u\partial_u\Phi-2(1-\Psi)\partial_u\xi^{r(0)}
			-2(1-\Phi)\partial_u\xi^u+2\Theta_B\partial_u\xi^{B(1)}\label{eq:dNgen}\\
			\delta m=&\xi^u\partial_u m-k(1-\Phi)(\xi^u+\xi^{r(0)})+\frac12\xi^{A(1)}D_A\Phi+K\partial_u\xi^{r(0)}\nonumber\\
			&-(1-\Psi)\partial_u\xi^{r(1)}+U_A\partial_u\xi^{A(1)}+\Theta_{B}\partial_{u}\xi^{B(2)}+2m\partial_u\xi^u\label{eq:dmgen}\\
			\delta \Psi=&\xi^u\partial_u \Psi-(1-\Psi)\partial_u\xi^u\label{eq:dEgen}\\
			\delta K=&\xi^u\partial_u K+K\partial_u\xi^u-2k(1-\Psi)(\xi^u+\xi^{r(0)})
			+\xi^{B(1)}(\partial_{B}\Psi-\Theta_B)\label{eq:dFgen}\\
			\delta \Omega=&\xi^u\partial_u \Omega\label{eq:dCgen}\\
			\delta C_{AB}=&\xi^u\partial_uC_{AB}+2(1+k)\gamma_{AB}(1+\Omega)\xi^{r(0)}
			+2k\xi^u(1+\Omega)\gamma_{AB}\nonumber\\
			&+(1+\Omega)\left(D_A\xi_B^{(1)}+D_B\xi_A^{(1)}\right)+\gamma_{AB}\xi^{C(1)}D_C \Omega \nonumber \\
			&+\Theta_AD_B\xi^u+\Theta_BD_A\xi^u\label{eq:dCABgen}\\
			\delta \Theta_A=&\xi^u\partial_u\Theta_A+\gamma_{AB}(1+\Omega)\partial_u\xi^{B(1)}+\Theta_A\partial_u\xi^u\label{eq:dAAgen}\\
			\delta U_A=&\xi^u\partial_u U_A+U_A\partial_u\xi^u+\Theta_A\xi^{r(0)}+2k\Theta_A(\xi^u+\xi^{r(0)})-(1-\Phi)D_A\xi^u\nonumber\\
			&-(1-\Psi)D_A\xi^{r(0)}+\xi^{B(1)}D_B\Theta_A+\Theta_BD_A\xi^{B(1)}+C_{AB}\partial_u\xi^{B(1)}\nonumber\\
			&+\gamma_{AB}(1+\Omega)\partial_u\xi^{B(2)}\label{eq:dUAgen} \\
			\delta N_A=&\xi^u\partial_u N_A+N_A\partial_u\xi^u+V^BD_BN_A+N_BD_AV^B-(1-2k)N_A\xi^{r(V)}\nonumber\\
			&+\xi^{B(1)}D_BU_A+U_BD_A\xi^{B(1)}+\xi^{B(2)}D_B\Theta_A+\Theta_BD_A\xi^{B(2)}\nonumber\\
			&-(1-\Psi)D_A\xi^{r(1)}+2mD_A\xi^u+2kU_A\left(\xi^{r(0)}+\xi^u-u\xi^{r(V)}\right)\nonumber\\
			&+2k\Theta_A\left(u^2\xi^{r(V)}-u(\xi^{r(0)}+\xi^u)+\xi^{r(1)}\right)+\Theta_A\xi^{r(1)}+C_{AB}\partial_u\xi^{B(2)}\nonumber\\
			&+KD_A\xi^{r(0)}+h_{AB}\partial_u\xi^{B(1)}\label{eq:dNAgen} \ .
		\end{align}
		\item Note that setting $\Phi=\Psi=K=\Omega=\Theta_A=0$ and $k=0$ we consistently recover the same results as for asymptotically flat spacetimes.
	\end{enumerate}
	
	\subsubsection*{New physical insight}
	
	A closer look to equations (\ref{eq:dNgen})-(\ref{eq:dUAgen}) reveals a much richer picture than in the flat case. Let us point out some of the more relevant features
	
	\begin{itemize}
		\item Terms with $g_{uu}\sim\mathcal{O}(r^{-1})$ and $g_{ur}\sim\mathcal{O}(r^{-1})$ are unavoidably generated for $k\neq0$, as can be seen from (\ref{eq:dmgen}) and (\ref{eq:dFgen}).
		\item For general setting and supertranslations, all the modes transform non-linearly except $\Omega$.
		\item While $\Psi$ and $\Omega$ receive at most contributions from themselves, $\Phi$, $K$, $\Theta_A$, $C_{AB}$ and $N_{A}$ feel the effect of other modes and $m$ can be affected by all the modes except $C_{AB}$ and $N_A$.
		\item Demanding the absence of terms $g_{ur}\sim\mathcal{O}(1)$ leads to $\partial_u\xi^u=0$, recovering the well-known result in the asymptotically flat case $\xi^u=f(z,\zb)\in\mathcal{C}^{\infty}(S^2)$. On the other hand, imposing $g_{uA}\sim\mathcal{O}(1)$ and/or $g_{uu}\sim\mathcal{O}(r^{-1})$ fixes the $u$-dependence of $\xi^u$. Otherwise, $\xi^u$ can depend arbitrarily on $u$ for a generic setting.
	\end{itemize}
	
	
	\subsubsection{Global Killing vectors}
	\label{gkvstrans}
	
	The goal of this section is to show how we recover the global Killing vectors (GKV) associated to translations consistently from the supertranslation diffeomorphisms (\ref{eq:Stranslac}).
	
	The global Killing vectors (GKV) are the solutions of the equation
	\begin{align}
		\mathcal{L}_{\xi}g_{\mu\nu}&=
		\xi^{\lambda}\partial_{\lambda}g_{\mu\nu}+g_{\nu\lambda}\partial_{\mu}\xi^{\lambda}
		+g_{\mu\lambda}\partial_{\nu}\xi^{\lambda}\overset{!}{=}0 \ .
	\end{align}
	Maximally symmetric spaces have the maximum number of GKV given by $d(d+1)/2$. In flat space, we obtain ten GKV, four associated to translations and six associated to rotations and boosts. Unperturbed FRW spaces are homogeneous and isotropic in the spatial components and, therefore, we obtain six GKV associated to the three spatial translations and three rotations. On the other hand, $\partial_t$ is no more a GKV but $\partial_{\eta}$ is a Conformal Killing Vector (CKV). 
	
	The large $r$-limit of (\ref{eq:Stranslac}) when $\Phi,\Omega,\Psi,\Theta_A\to0$ is given by
	\begin{align}
		\xi=&\xi^u(u,z,\zb)\partial_u+\frac{1}{2(1+k)}\left[D_AD^A\xi^u-2k\xi^u+\mathcal{O}(r^{-1})\right]\partial_r\\
		&+\left[-\frac{1}{r}D^B\xi^u
		+\mathcal{O}(r^{-2})\right]\partial_B+\mathcal{O}(\Phi,\Omega,\Psi,\Theta) \ .
		\label{eq:StranslaclimitNCFE0}
	\end{align}
	
	More concretely, if we only allow for $\Lie{\xi}{g_{ur}}=\Op(r^{-1})$, $\xi^{u}=f(z,\zb)+\mathcal{O}(\Psi)$ in the limit $\Psi\to0$ and we obtain:
	\begin{align}
		\xi=&f(z,\zb)\partial_u+\frac{1}{2(1+k)}\left[D_AD^Af(z,\zb)-2kf(z,\zb)+\mathcal{O}(r^{-1})\right]\partial_r\\
		&+\left[-\frac{1}{r}D^B f(z,\zb)
		+\mathcal{O}(r^{-2})\right]\partial_B+\mathcal{O}(\Phi,\Omega,\Psi,\Theta) \ ,
		\label{eq:StranslaclimitNCFE0ur-1}
	\end{align}
	whose action is equivalent to the following coordinate transformations:
	\begin{align}
		u\to u+f \ , \ \ \ r\to r+\frac{1}{2(1+k)}\left(D^AD_A f-2kf\right)   \\
		z\to z-\frac{1}{r}D^zf \ , \ \ \ \bar{z}\to \bar{z}-\frac{1}{r}D^{\bar{z}}f
	\end{align}
	
	Using the following convention \cite{Strominger:2017zoo} for the $l=0,1$ spherical armonics 
	\begin{align}
		Y^{0}_{0}=1 \ , \ \ Y^{1}_{1}=\frac{z}{1+z\bar{z}} \ , \ \ Y^{0}_{1}=\frac{1-z\bar{z}}{1+z\bar{z}} \ , \ \ Y^{-1}_{1}=\frac{\bar{z}}{1+z\bar{z}} \ ,
	\end{align}
	we aim to recover the unperturbed FRW GKVs generating the spatial translations as some linear combinations of $\xi(Y^{0}_{0})$, $\xi(Y^{1}_{1})$, $\xi(Y^{0}_{1})$ and $\xi(Y^{-1}_{1})$
	\begin{align}
		\xi(Y^{0}_{0})&=\partial_{u}-\frac{k}{(1+k)}\partial_{r}=\partial_{\eta}-\frac{(1+2k)}{(1+k)}\partial_r\\
		\xi(Y^{1}_{1})&=\frac{z}{1+z\zb}\left( \partial_u-\partial_r \right)+\frac{1}{r} \left( \frac{z^2}{2}\partial_z-\frac{1}{2}\partial_{\zb} \right)\\
		\xi(Y^{0}_{1})&=\frac{1-z\zb}{1+z\zb}(\partial_u-\partial_r)+\frac{1}{r} (z\partial_z +\zb\partial_{\zb})\\
		\xi(Y^{-1}_{1})&=\frac{\zb}{1+z\zb}(\partial_u-\partial_r)+\frac{1}{r} \left( -\frac{1}{2}\partial_z+\frac{\zb^2}{2}\partial_{\zb}\right) \ .
	\end{align}
	
	We can write them in terms of the Cartesian generators $X_i=\partial_{x^{i}}$ as:
	\begin{equation}
		\xi(Y^0_1)=-X_3 \ , \ \ \ \xi(Y^1_1)=-\frac{1}{2}(X^1+iX^2) \ , \ \ \ \xi(Y^{-1}_1)=-\frac{1}{2}(X^1-iX^2) \ ,
		\label{eq:rotations}
	\end{equation}
	obtaining exactly the same result as in flat space \cite{Strominger:2017zoo}.
	
	Consistently, we do not obtain the time translation generator from $\xi(Y^0_0)$, due to the fact that it is not a GKV in spatially flat FRW but a CKV. Therefore, there is a linearly independent (and so unavoidable) correction term which corresponds to an spatial dilatation $D$ pondered by the inverse of the radius
	\begin{equation}
		\xi(Y^0_0)=\frac{1}{(1+k)}\left[\partial_{\eta}-\frac{k}{r}x_i\partial_{x_i}\right]=\frac{1}{(1+k)}\left[T_{\text{Conf}}-\frac{k}{r}D\right] \ .
		\label{eq:ckvtime}
	\end{equation}
	
	\paragraph{Remarks}
	
	\begin{itemize}
		\item In case of considering flat BMS supertranslations (as in \cite{Kehagias:2016zry}), instead of the asymptotically spatially flat FRW supertranslations that we study, the relations (\ref{eq:rotations}) would be analogously verified, but (\ref{eq:ckvtime}) would be replaced by $\xi(Y^0_0)=T_{\text{Conf}}$ being in line with the fact that pure spatially flat FRW is conformal to Minkowski after replacing $t$ by $\eta$. However, we observe from the above discussion that we do not recover the correct global isometry group of FRW from flat BMS supertranslations. Therefore, flat BMS are not consistent asymptotic symmetries in our cosmological setting. 
		\item If the coefficients $\Phi,\Omega,\Psi,\Theta_A$ are non zero, they survive at infinity and we should recover the GKV of the corresponding perturbed spatially flat FRW. Nevertheless, such spaces generally do not have GKV. Although, if $\Phi,\Omega,\Psi,\Theta_A<<1$ \footnote{As physically expected, otherwise the perturbations would spoil the observed homogeneity and isotropy at large $r$ scales in our universe.}, then one can expand them in series and we obtain the previous results as a first approximation.
		\item The same results follow for the general case (\ref{eq:StranslaclimitNCFE0}) from the $u$-independent and $l=0,1$ spherical harmonic modes of $\xi^u$.
	\end{itemize}

	
	\subsection{Superrotations}
	\label{sec_asfFLRWsrot}
	
	Next, we allow also for superrotations. The ansatz for infinitesimal diffeomorphisms generating superrotations is similar to the supertranslations, but includes an $\Op(r)$ contribution in the $r$ component and an $\Op(r^0)$ contribution in the angular components \footnote{The fact that $\xi^u$ has to be $r$-independent follows from (\ref{eq:bondigauge1}).}
	\begin{align}
		\xi_R=\xi^u(u,x^A)\partial_u+\left(r\xi^{r(V)}+\sum\limits_{n=0}^{\infty}\frac{\xi^{r(n)}}{r^n}\right)\partial_r+\left(V^A+\sum\limits_{n=1}^{\infty}\frac{\xi^{A(n)}}{r^n}\right)\partial_A \ .
		\label{eq:superrot_ansatz}
	\end{align}

	
	\subsubsection{General case}
	
	The first step is to fix the Bondi gauge. $0=\mathcal{L}_{\xi}g_{rr}$ is automatically verified by (\ref{eq:superrot_ansatz}) and $0=\mathcal{L}_{\xi}g_{rA}$ provides us with the Lie derivative (\ref{eq:liesrotrA}) and same conditions (\ref{eq:xi1}) and (\ref{eq:xi2}) as before. The last gauge condition we need to impose is (\ref{eq:detBondisutra}). Following the same steps as in section \ref{asflatFLRWstrans}, we obtain (\ref{eq:explicitbdgauge}), which, together with $C^A_A=0$ and (\ref{eq:LieAB11srot}), leads us to
	\begin{align}
		\xi^{r(0)}=&\frac{1}{1+k}\left[-\frac{1}{2(1+\Omega)}D_A((1+\Omega)\xi^{A(1)})-\frac{1}{2(1+\Omega)}\Theta^AD_A\xi^u+ku\xi^{r(V)}-k\xi^u\right] \ ,
		\label{eq:coeffstransr0s} \\
		\xi^{r(1)}=&\frac{1}{2(1+k)(1+\Omega)}\left[\frac{C^A_B\Theta_AD^B\xi^u}{1+\Omega}-2k(1+\Omega)\left(u^2\xi^{r(V)}-u\xi^{r(0)}-u\xi^u\right)\right. \nonumber\\
		&\left.-D_A\left((1+\Omega)\xi^{A(2)}\right)-U^AD_A\xi^u\right] \ .
		\label{eq:coeffstransr1s}
	\end{align}
	
	Contrary to the supertranslations, the Lie derivatives (appendix \ref{liedersrot}) do not close in the $r$ expansion (\ref{eq:asymptpertspatflatFLRW}). In order to be consistent with (\ref{eq:asymptpertspatflatFLRW}), the remaining falloff conditions we require are
	\begin{align}
		\Lie{\xi}{g_{uu}}=&\Op(1),\quad \Lie{\xi}{g_{ur}}=\Op(1),\quad \Lie{\xi}{g_{uA}}=\Op(r),\quad \\ &\Lie{\xi}{g_{AB}}=(1+\Omega+\delta \Omega)\gamma_{AB}+\Op(r) \ .
	\end{align}
	
	From the Lie derivatives (\ref{eq:uusrot}) and (\ref{eq:uAsrot}) we get the additional requirements
	\begin{align}
		\partial_uV^A=\partial_u\xi^{r(V)}=0 \ ,
	\end{align}
	and, from $F_{AB}$ in (\ref{eq:LieAB11srot}), we have to restrict $V^C$ to be conformal Killing vectors on the sphere
	\begin{align}
		D_AV_B+D_BV_A=\gamma_{AB}D_CV^C \ .
	\end{align}
	
	
	\subsubsection*{Summary of results}
	
	\begin{enumerate}
		\item General superrotations have the form:
		\begin{align}
			\xi=&\xi^u(u,z,\zb)\partial_u+\left[r\xi^{r(V)}(z,\zb)+\xi^{r(0)}+\frac{1}{r}\xi^{r(1)}\right]\partial_r \nonumber \\
			&+\left[V^{B}(z,\zb)+\frac1r\xi^{B(1)}+\frac{1}{r^2}\xi^{B(2)}\right]\partial_B \ .
			\label{eq:Srotacc}
		\end{align}
		\item There are three free parameters, namely $\xi^u(u,z,\zb)$, $V^{B}(z,\zb)$ and $\xi^{r(V)}(z,\zb)$. 
		\item The other coefficients in \eqref{eq:Srotacc} are given by (\ref{eq:coeffstransr0s}), (\ref{eq:coeffstransr1s}), (\ref{eq:xi1}) and (\ref{eq:xi2}).
		\item From the Lie derivatives in appendix \ref{liedersrot}, we observe that the action of \eqref{eq:Srotacc} satisfies the falloff conditions (\ref{eq:asymptpertspatflatFLRW}) automatically, as long as $V^A$ are CKV on the sphere, and induces the following transformations in the asymptotic data:
		\begin{align}
			\delta\Phi=&V^AD_A\Phi+\xi^u\partial_u\Phi-2(1-\Psi)\partial_u\xi^{r(0)}-2k(1-\Phi)\xi^{r(V)}\nonumber\\
			&-2(1-\Phi)\partial_u\xi^u+2\Theta_A\partial_u\xi^{A(1)}\label{eq:dNgenrot}\\
			\delta m=&\xi^u\partial_u m-k(1-\Phi)\xi^u-\left((1-2k)m-ku(1-\Phi)\right)\xi^{r(V)}\nonumber\\
			&-k(1-\Phi)\xi^{r(0)}+V^AD_Am+\frac12\xi^{A(1)}D_A\Phi+K\partial_u\xi^{r(0)}\nonumber\\
			&-(1-\Psi)\partial_u\xi^{r(1)}+m\partial_u\xi^u+U_A\partial_u\xi^{A(1)}+\Theta_A\partial_u\xi^{A(2)}\label{eq:dmgenrot}\\
			\delta \Psi=&V^A\partial_A\Psi+\xi^u\partial_u\Psi-(1+2k)(1-\Psi)\xi^{r(V)}-(1-\Psi)\partial_u\xi^u\label{eq:dEgenrot}\\
			\delta K=&\xi^u\partial_u K+V^AD_AK+K\partial_u\xi^u+\xi^{A(1)}D_A\Psi-\Theta_A\xi^{A(1)}\nonumber\\
			&+2k(1-\Psi)\left(u\xi^{r(V)}-\xi^u-\xi^{r(0)}\right)+2kK\xi^{r(V)}\label{eq:dFgenrot}\\
			\delta \Omega=&V^CD_C\Omega+\xi^u\partial_u\Omega+2(1+k)\xi^{r(V)}+(1+\Omega)D_AV^A\label{eq:dCgenrot}\\
			\delta C_{AB}=&\xi^u\partial_uC_{AB}+V^CD_C C_{AB}+C_{AC}D_BV^C+C_{BC}D_AV^C\nonumber\\
			&+2(1+\Omega)\gamma_{AB}((1+k)\xi^{r(0)}-ku\xi^{r(V)}+k\xi^u)+\gamma_{AB}\xi^{C(1)}D_C \Omega\nonumber\\
			&+(1+\Omega)(D_A\xi_B^{(1)}+D_B\xi_A^{(1)})+\Theta_AD_B\xi^u+\Theta_BD_A\xi^u \nonumber\\
			&+(1+2k)C_{AB}\xi^{r(V)}\label{eq:dCABgenrot}\\
			\delta \Theta_A=&V^BD_B\Theta_A+\xi^u\partial_u\Theta_A+(1+2k)\Theta_A\xi^{r(V)}+\Theta_BD_AV^B\nonumber\\
			&-(1-\Psi)\partial_A\xi^{r(V)}+\Theta_A\partial_u\xi^u+(1+\Omega)\partial_u\xi^{(1)}_A\label{eq:dAAgenrot}\\
			\delta U_A=&(2k\Theta_A+\partial_u U_A)\xi^u+(1+2k)\Theta_A\xi^{r(0)}+2k\xi^{r(V)}(U_A-u\Theta_A)\nonumber\\
			&+V^BD_BU_A+\xi^{B(1)}D_B\Theta_A+\Theta_BD_A\xi^{B(1)}+U_BD_AV^B\nonumber\\
			&-(1-\Psi)D_A\xi^{r(0)}+KD_A\xi^{r(V)}-(1-\Phi)D_A\xi^u+U_A\partial_u\xi^u\nonumber\\
			&+C_{AB}\partial_u\xi^{B(1)}+(1+\Omega)\partial_u\xi_A^{(2)}\label{eq:dUAgenrot}\\
			\delta N_A=&\xi^u\partial_u N_A+N_A\partial_u\xi^u+V^BD_BN_A+N_BD_AV^B-(1-2k)N_A\xi^{r(V)}\nonumber\\
			&+\xi^{B(1)}D_BU_A+U_BD_A\xi^{B(1)}+\xi^{B(2)}D_B\Theta_A+\Theta_BD_A\xi^{B(2)}+KD_A\xi^{r(0)}\nonumber\\
			&-(1-\Psi)D_A\xi^{r(1)}+2mD_A\xi^u+2kU_A\left(\xi^{r(0)}+\xi^u-u\xi^{r(V)}\right)\nonumber\\
			&+2k\Theta_A\left(u^2\xi^{r(V)}-u(\xi^{r(0)}+\xi^u)+\xi^{r(1)}\right)+\Theta_A\xi^{r(1)}+C_{AB}\partial_u\xi^{B(2)}\nonumber\\
			&+h_{AB}\partial_u\xi^{B(1)}\label{eq:dNAgenrot}
		\end{align}
		\item Note that setting $\Phi=\Psi=K=\Omega=\Theta_A=0$ and $k=0$ we consistently recover the same results as for asymptotically flat spacetimes. Furthermore, imposing $\xi^{r(V)}=V^B=0$ leads us to the analysis of section \ref{asflatFLRWstrans}. 
	\end{enumerate}
	
	
	\subsubsection*{New physical insight}
	
	Equations (\ref{eq:dNgenrot})-(\ref{eq:dNAgenrot}) show a richer picture than allowing only for supertranslations. Let us briefly mention some of the more relevant features:
	
	\begin{itemize}
		\item For general setting and superrotations, all the modes transform non-linearly.
		\item Demanding the absence of terms $g_{ur}\sim\mathcal{O}(1)$ fixes $\xi^{r(V)}$ to be
		\begin{align}
			\xi^{r(V)}=-\frac{1}{(1+2k)}\partial_u\xi^u \ .
		\end{align}
		This means that $\xi^u$ is at most linear in $u$ as in the asymptotically flat case. Besides, imposing $g_{uA}\sim\mathcal{O}(1)$ and/or $g_{uu}\sim\mathcal{O}(r^{-1})$ fixes the $u$-dependence of $\xi^u$. Apart from that, $\xi^u$ can depend arbitrarily on $u$ for a generic setting.
		\item For $k\neq0$, $\xi^{r(V)}$ generates unavoidable contributions for all the modes except $C_{AB}$. Remarkably, $m$ and $K$ become ``dynamical'' through the $u$-dependent term $ku\xi^{r(V)}$.
		\item $V^A$ together with $\xi^{r(V)}$ generate an inevitable term in $\delta\Theta_A$.
		\item For $k\neq0$, $\xi^{r(V)}$ enters directly the subleading $\xi^{r(0)}$ and $\xi^{r(1)}$ components of the asymptotic superrotation diffeomorphisms.
	\end{itemize}
	
	
	\subsubsection*{Strong Bondi gauge}
	
	In the asymptotically flat case, the determinant (\ref{eq:explicitbdgauge}) is usually required to be exactly $\det(g)$ \cite{Hamada:2017gdg}, which is a stronger requirement than just the Bondi gauge (\ref{eq:detBondisutra}) \cite{Ruzziconi:2020cjt}. Following the same path would lead to demand that $F^A_A=0$, which gives the requirement
	\begin{align}
		V^CD_{C} \Omega+2(1+k)\xi^{r(V)}+\xi^u\partial_u\Omega+(1+\Omega)D_AV^A=0 \ .
		\label{eq:sbg}
	\end{align}
	This condition fixes $\xi^{r(V)}$ \footnote{When the parameter $\xi^{r(V)}$ is free, it represents, to leading order in $r$ ($\xi\propto \xi^{r(V)}r\partial_r$), angle dependent dilatations. They might be related to the superdilations in \cite{Donnay:2020fof}.} in relation to $V^A$ and $\xi^u$
	\begin{align}
		\xi^{r(V)}=-\frac{1}{2(1+k)}[\xi^u\partial_u\Omega+D_A((1+\Omega)V^A)] \ ,
		\label{eq:xirvfix}
	\end{align}
	which now remain independent as the unique parameters in the asymptotic symmetry group.
	
	Note that we could have required the strong Bondi gauge already for the supertranslations $\xi^{r(V)}=V^A=0$. This would force $\Omega$ to be independent of $u$, such that it would not be a dynamical field in the asymptotic expansion but a coordinate transformation. 
	
	Let us point out that replacing (\ref{eq:xirvfix}) in (\ref{eq:LieAB11srot}) leads to
	\begin{equation}
		F_{AB}=(1+\Omega)(D_AV_B+D_BV_A-\gamma_{AB}D_CV^C) \ ,
		\label{eq:Cdiffeo}
	\end{equation}
	meaning that $V^A$ must be a CKV on the sphere if we require $\delta\gamma_{AB}=0$, identically to the situation encountered in the asymptotically flat case (section \ref{aflatsrot}) \footnote{Alternatively, one could allow for general diffeomorphisms on $S^2$ but then the round metric $\gamma_{AB}$ should be replaced by a general $q_{AB}(u,z,\zb)$ in (\ref{eq:asymptpertspatflatFLRW}) and the entire analysis repeated. Note that, for a $u$-dependent $\Omega$, it is clear from (\ref{eq:Cdiffeo}) that an $u$-dependent $q_{AB}$ would be generated.}.

	To summarize, imposing the strong Bondi gauge leads to $\mathcal{L}_{\xi}g_{AB}\sim\mathcal{O}(r)$ causing $\delta \Omega=\delta \gamma_{AB}=0$. In the case of supertranslations, it forces $\partial_u\Omega=0$ and for superrotations $\Omega$ can still be a dynamical coefficient but severely restricting the $u$-dependence of $\xi^{u}$ through $\partial_u\xi^{r(V)}=\partial_{u}V^A=0$ and (\ref{eq:xirvfix}). 
	
	
	\subsubsection{Particular cases}
	\label{partcases}
	
	We notice that the inclusion of dynamical $\Omega$ is in tension with the strong Bondi gauge, potentially causing incompatibilities with many interesting cosmological settings and opening several possibilities:
	
	\begin{enumerate}
		\item \textit{Allow for arbitrary $\Omega(u,z,\zb)$.}
		
		The application of the strong Bondi gauge is not appropriate because then $\delta\Omega=0$, preventing us from analyzing how it transforms under the asymptotic diffeomorphisms.
		
		\item \textit{Restrict to metrics with $\partial_u\Omega=0$.}
		
		$\Omega$ is not a dynamical cosmological mode, $\xi^u$ is unconstrained and the strong Bondi gauge is applicable. The normal Bondi gauge allows for $z,\zb$-dependent contributions to $\Omega$, while the strong Bondi gauge freezes any change in $\Omega$.

		\item \textit{Limit to $\xi^{r(V)}=0$.}
		
		Taking into account that boost generators are not GKV for FRW (c.f. section \ref{gkvsrot}), it is not necessary to add the $r\xi^{r(V)}$ term in (\ref{eq:superrot_ansatz}). In fact, it is possible to restrict to this subclass of diffeomorphisms, being still consistent with the obtainment of the GKVs of FRW.

		\item \textit{Perform again the analysis to allow for general diffeomorphisms on $S^2$.}
		
		It would permit us to describe a richer variety of cosmological perturbations to FRW at leading order, including anisotropies of the form $r^2[\lambda(u,z,\zb) dz^2+\beta(u,z,\zb) d\zb^2]\subset ds^2$, obviously incompatible with the strong Bondi gauge. We leave this for future research due to its length and technical difficulty. 
	\end{enumerate}
	
	\paragraph{Remark:}
	It is important to clarify that allowing for a dynamical $\Omega$ provides us with a third scalar mode, such that there is still gauge freedom. From the perspective of analyzing cosmological perturbations, we could (and should) fix totally the gauge and restrict to metrics with $\Omega=0$, but it is still fruitful to leave this coefficient free from the perspective of exploring off-shell the most general asymptotic transformations between asymptotically spatially flat FRW spacetimes. In section \ref{sect_cosmopert}, we will come back to this point in more detail.

	
	\subsubsection{Global Killing vectors}
	\label{gkvsrot}
	
	In this section, we consistently obtain the global Killing vectors, associated to rotations, from the superrotation diffeomorphisms (\ref{eq:Srotacc}). In contrast to the asymptotically flat case, the boosts are no longer global Killing vectors.
	
	The first step is to derive $\xi$ for the global CKV on $S^2$ in the limit $\Phi,\Omega,\Psi,\Theta_A\to0$. For the sake of simplicity, we adopt the strong Bondi gauge. We find from (\ref{eq:xirvfix}), (\ref{eq:xi1}) and (\ref{eq:ursrot}) that 
	\begin{align}
		\xi^{r(V)}=&-\frac{1}{2(1+k)}D_AV^A \ , \\
		\xi^{A(1)}=&-D^A\xi^u \ .
	\end{align}
	To fix $\xi^u$ we have to either make the leading contribution to \eqref{eq:ursrot} or \eqref{eq:uAsrot} vanish. Note that for $\xi^{r(V)}\neq0$ there is no way to make both terms vanish at the same time. That means we will always generate a $\Psi$ or $\Theta_A$ term in the metric. Comparing this to the situation in asymptotically flat space \cite{Strominger:2017zoo}, we see that the transformations with $\xi^{r(V)}\neq0$ include the boosts. As a consequence, our observation that for those transformations we cannot make all the leading terms in the Lie derivative vanish reflects the fact that boosts are no longer GKVs of FRW.
	
	The two different choices for $\xi^u$ that make $\delta\Psi$ or $\delta\Theta_A$ vanish respectively are given by
	\begin{align}
		\delta\Psi=0:&\quad\Rightarrow \xi_{\Psi}^{u}=\frac{u}{2}\frac{(1+2k)}{(1+k)}D_AV^A+f(x^A)\\
		\delta\Theta_A=0:&\quad\Rightarrow \xi_{\Theta}^{u}=\frac{u}{2}\frac{1}{1+k}D_AV^A+f(x^A) \ .
	\end{align}
	
	Setting $f\to0$ and using these equations and the fact that $D_AD^AD_BV^B=-2D_BV^B$ for CKV on the sphere, (\ref{eq:Srotacc}) becomes
	\begin{align}
		\xi_{\Psi}=&\frac{u}{2}\frac{(1+2k)}{(1+k)}D_AV^A\partial_u-\frac{1}{2(1+k)^2}[r(1+k)+u(1+4k+2k^2)]D_AV^A\partial_r \nonumber \\
		&+\left(V^A-\frac{u}{2r}\frac{(1+2k)}{(1+k)}D^AD_BV^B\right)\partial_A\label{eq:boost_psi_bondi}\\
		\xi_{\Theta}=&\frac{u}{2}\frac{1}{(1+k)}D_AV^A\partial_u-\frac{1}{2(1+k)^2}[r(1+k)+u(1+2k)]D_AV^A\partial_r \nonumber \\
		&+\left(V^A-\frac{u}{2r}\frac{1}{(1+k)}D^AD_BV^B\right)\partial_A \ . \label{eq:boost_theta_bondi}
	\end{align}
	
	It is straightforward to check that the choices
	\begin{align}
		V^z=iz \ \ \ \ \ \ \ \ \ \ & \ \ \ \ \ \ \ \ V^{\zb} =-i\zb  \nonumber\\
		V^z=\frac{i}{2}(z^2-1)   & \ \ \ \ \ \ \ \ V^{\zb} =\frac{i}{2}(1-\zb^2)  \nonumber\\
		V^z=\frac{1}{2}(1+z^2)   & \ \ \ \ \ \ \ \ V^{\zb} =\frac{1}{2}(1+\zb^2) \label{eq:GKVrotations}
	\end{align}
	verify $D_AV^A=0$ which means $\xi^{r(V)}=0$ and correspond respectively to the rotation generators $J_{12}$, $J_{23}$ and $J_{31}$, where $J_{ij}=x^i\partial_j-x^j\partial_i$ in Cartesian coordinates.
	
	Moving on to CKVs with $D_AV^A\neq0$, we can consider the choices
	\begin{align}
		V^z&=\frac12(1-z^2) && V^{\zb}=\frac12(1-\zb^2)\\
		V^z&=\frac{i}{2}(1+z^2) && V^{\zb}=-\frac{i}{2}(1+\zb^2)\\
		V^z&=-z && V^{\zb}=-\zb \ .\label{eq:boost_z}
	\end{align}
	In the asymptotically flat case these correspond to the boosts in $x,y$ and $z$ direction respectively. To see how a flat boost would look like in our case we plug \eqref{eq:boost_z} into \eqref{eq:boost_psi_bondi} and \eqref{eq:boost_theta_bondi}. After transforming the result into cartesian coordinates we discover that these transformations can be written in terms of a conformal boost term perturbed by a superposition of deformed conformal transformations:
	\begin{align}
		\xi^{(i)}_{\Psi}=&\frac{1}{(1+k)^2}\left\{B_i+\frac{k}{r}\left[\left(1+k-(3+2k)\frac{\eta}{r}\right)K_i\right.\right.\\
		&\left.\left.+\left((6+5k)\frac{\eta}{r}-(5k+4)\right)x_iD+x_i\eta T_c\right]\right\}\\
		\xi^{(i)}_{\Theta}=&\frac{1}{(1+k)^2}\left\{B_i+\frac{k}{r}\left[-\left(1+k+\frac{\eta}{r}\right)K_i+\left(3\frac{\eta}{r}+k\right)x_iD+x_i\eta T_c\right]\right\} \ ,
	\end{align}
	
	where the boosts $B_i$, special conformal transformations $K_i$, dilatation $D$ and conformal time translation $T_c$ are given by:
	\begin{align}
		B_i=&\eta\partial_i+x_i\partial_{\eta}; && K_i=2x_i x^j\partial_j-r^2\partial_i; && D=x^i\partial_i; && T_c=\partial_{\eta} \ .
	\end{align}
	
	Analogous remarks to those at the end of section \ref{gkvstrans} apply here.
	
	
	\section{Examples}
	\label{sect_examples}
	To get a feeling for the physical meaning of the transformations we considered in the previous section, we will now apply them to some example spacetimes. We start with a pure FRW universe and realize that the asymptotic transformations can be classified into orbits according to the leading order terms they generate. 
	We move on to describe the transformations of a FRW universe with a non-vanishing Bondi mass.
	Finally, we consider cosmological perturbations and their transformation behaviour.
	
	
	\subsection{Pure FRW}
	\label{subsec_pureFRW}
	
	The asymptotic data and diffeomorphisms are considerably simplified starting from pure FRW
	\begin{align}
		\delta\Phi_{FRW}=&-2\partial_u\xi^{r(0)}-2k\xi^{r(V)}-2\partial_u\xi^u\label{eq:dNgenrotpure}\\
		\delta m_{FRW}=&-k\xi^u+ku\xi^{r(V)}-k\xi^{r(0)}-\partial_u\xi^{r(1)}\label{eq:dmgenrotpure}\\
		\delta \Psi_{FRW}=&-(1+2k)\xi^{r(V)}-\partial_u\xi^u\label{eq:dEgenrotpure}\\
		\delta K_{FRW}=&2k\left(u\xi^{r(V)}-\xi^u-\xi^{r(0)}\right)\label{eq:dFgenrotpure}\\
		\delta \Omega_{FRW}=&2(1+k)\xi^{r(V)}+D_AV^A\label{eq:dCgenrotpure}\\
		\delta C_{AB\ FRW}=&2\gamma_{AB}((1+k)\xi^{r(0)}-ku\xi^{r(V)}+k\xi^u)+(D_A\xi_B^{(1)}+D_B\xi_A^{(1)})\label{eq:dCABgenrotpure}\\
		\delta \Theta_{A\ FRW}=&-D_A\xi^{r(V)}+\partial_u\xi^{(1)}_A\label{eq:dAAgenrotpure}\\
		\delta U_{A\ FRW}=&-D_A\xi^{r(0)}-D_A\xi^u\label{eq:dUAgenrotpure}\\
		\delta N_{A\ FRW}=&-D_A\xi^{r(1)}\label{eq:dNAgenrotpure} \ .
	\end{align}
	
	\begin{align}
		\xi^{A(1)}_{FRW}=&-D^A\xi^u; && \xi^{r(0)}_{FRW}=\frac{1}{1+k}\left[\frac{1}{2}\left(D_AD^A-2k\right)\xi^u+ku\xi^{r(V)}\right]  \\
		\xi^{A(2)}_{FRW}=&0; && \xi^{r(1)}_{FRW}=\frac{ku}{(1+k)^2}\left[\frac{1}{2}\left(D_AD^A+2\right)\xi^u-u\xi^{r(V)}\right] \ .
	\end{align}
	
	Note that, for general transformations, all the asymptotic data can be generated. Particularly, this means the generation of cosmological modes out of the vacuum. This is, however, to be expected as the asymptotic transformations are in general no GKVs of FRW as we described in the previous section. An observer at null infinity undergoing an asymptotic transformation that is not a GKV will therefore see a different spacetime that is slightly deformed compared to the original FRW. In contrast to an asymptotically flat spacetime, a FRW universe is filled with a homogeneous and isotropic hydrodynamic fluid everywhere. The observer at null infinity will therefore experience the deformations of the universe as perturbations in the distribution and flux of the fluid which is why all the asymptotic data can be generated from the FRW vacuum by a general transformation. This effect does not exist in the asymptotically flat case.
	
	
	\subsubsection{Orbits of transformations}
	\label{subsubsect_orbits}
	
	We classify the general transformations into orbits according to which leading order terms vanish. This is equivalent to enforcing a certain extra fall off behaviour in the metric.
	\begin{enumerate}
		\item $\delta \Psi=0   \Longrightarrow \xi^{u}=-u(1+2k)\xi^{r(V)}+f(z,\zb)+\text{const}$
		\item $\delta \Phi=0   \Longrightarrow k\xi^{r(V)}=-\frac{(D_AD^A+2)}{2(1+k)}\partial_u\xi^u $
		\item $\delta \Theta_A=0   \Longrightarrow \xi^{u}=-u\xi^{r(V)}+f(z,\zb)+\text{const} $
		\item $\delta \Omega=0   \Longrightarrow \xi^{r(V)}=-\frac{1}{2(1+k)}D_AV^A$  \qquad  (strong Bondi gauge) \ .
	\end{enumerate}
	
	In the upcoming analysis we take into account that $k\in\mathbb{Q}\setminus[(-\infty,0)\cup(1,\infty)]$, that $D^2=D_AD^A$ has as unique eigenfunctions on $S^2$ the trivial function and the spherical harmonics $Y^l_m$ with eigenvalues $-l(l+1)$ , $l=0,1,2 ...$ and that $V^A$ being a CKV implies that $D^2D_AV^A=-2D_AV^A$, such that $D_AV^A=\sum_i a_i^V Y^1_m$. 
	
	Compatibility by pairs:
	
	\begin{itemize}
		\item $\delta\Omega=0$ is compatible with all the others, as it is the only one involving the parameter $V^A$.
		\item $\delta\Psi=0$ is only compatible with $\delta\Theta_A=0$ if $k=0$ and/or $\xi^{r(V)}=0$ 
		\item $\delta\Psi=0$ is compatible with $\delta\Phi=0$ if $\xi^{r(V)}=0$ or if $\xi^{r(V)}=\sum_i b^i Y^1_m\neq0$ and $k=0$.
		\item $\delta\Phi=0$ is compatible with $\delta\Theta_A=0$ if $\xi^{r(V)}=0$, if $\xi^{r(V)}=\sum_i c^i Y^1_m\neq0$ and $k=0$, or if $\xi^{r(V)}\propto Y^1_0$ and $k=\frac12$ (free scalar field) \footnote{It is interesting to note that another solution exists outside our regime, that is $\xi^{r(V)}\propto Y^1_0$ and $k=-1$ (scalar field with vanishing kinetic energy).}.
	\end{itemize}
	
	We observe the following interesting facts:
	
	\begin{itemize}
		\item If $\xi^{r(V)}=0$, then the first three conditions are verified together and can be extended to also include $\delta\Omega=0$ in case that $V^A$ are rotations. 
		
		\item For $k=0$, we recover the expected results for flat space. The four conditions are fulfilled simultaneously in the so-called strong Bondi gauge. If one allows for $\delta\Omega\neq0$, the other three conditions can only be verified if $\xi^{r(V)}=0$ or it consists of any linear combination of $l=1$ spherical harmonics. Otherwise, at most $\delta\Psi=0$ and $\delta\Theta_A=0$ are compatible.
		
		\item In case that $k\neq0$, the unique possibility to compatibilize more than two conditions is the previously discussed $\xi^{r(V)}=0$.
		
		\item Among the first three conditions, the only one which allows for arbitrary $u$-dependence on $\xi^u$ is $\delta\Phi=0$ in case that $k\xi^{r(V)}=0$ and at cost that its angular part is restricted to combinations of $Y^1_m$. For $k\neq0$, $\delta\Omega=0$ would only be possible for $V^A$ being rotations, whereas for $k=0$ it would be unconstrained.
	\end{itemize}

	\subsubsection{Particular transformations}
	
	Next, we show the non-vanishing contributions of some simple subsets of transformations to the asymptotic data.
	
	\begin{itemize}
		\item \textit{$\xi^{r(V)}=V^A=0$ and $\xi^u(u,z,\zb)=f=\text{const}$}
		\begin{align}
			\delta m=&-\frac{k(k+2)}{(1+k)^2}f; && \delta K=-\frac{2k}{1+k}f.
		\end{align}
		
		\item \textit{$\xi^{r(V)}=V^A=0$ and $\xi^u(u,z,\zb)=f(z,\bar{z})$}
		\begin{align}
			\delta m=&-\frac{k(k+2)}{2(1+k)^2}[D_AD^A+2]f(z,\bar{z}); && \delta C_{zz}=-2D_zD_zf(z,\bar{z});\\
			\delta K=&-\frac{k}{1+k}[D_AD^A+2]f(z,\bar{z}); && \delta C_{\bar{z}\bar{z}}=-2D_{\bar{z}}D_{\bar{z}}f(z,\bar{z});\\
			\delta U_A=&-\frac{1}{2(1+k)}D_A[D_CD^C+2]f(z,\bar{z});\\
			\delta N_A=&-\frac{ku}{2(1+k)^2}D_A[D_CD^C+2]f(z,\bar{z}) \ .
		\end{align}
		
		Notice that all variations vanish in the case $f(z,\bar{z})\propto Y^1_m$, which consistently correspond to the spatial translations analyzed in section \ref{gkvstrans}. Besides, $\delta m$, $\delta K$ and $\delta N_A$ are generated for $k\neq0$. 
		
		\item \textit{$\xi^{r(V)}=V^A=0$ and $\xi^u(u,z,\zb)=f(u)$}
		\begin{align}
			\delta\Phi=&-\frac{2}{1+k}\partial_uf; && \delta\Psi=-\partial_uf;\\
			\delta m=&-\frac{k}{(1+k)^2}[(k+2)+u\partial_u]f; && \delta K=-\frac{2k}{(1+k)}f \ .
		\end{align}
		
		\item \textit{$\xi^{r(V)}=0$, $D_AV^A=0$ and $\xi^u(u,z,\zb)=f(u)Y^1_m$}
		\begin{align}
			\delta\Psi=&-\partial_u\xi^u; && \delta\Theta_A=-D_A\partial_u\xi^u \ .
		\end{align}
		
		\item \textit{$\xi^u=V^A=0$ and $\xi^{r(V)}(z,\zb)=g$}
		\begin{align}
			\delta \Phi=&-\frac{2k(2+k)}{1+k}g; && \delta \Psi=-(1+2k)g;\\
			\delta m=&\frac{k(k+3)}{(1+k)^2}ug; && \delta K=\frac{2k}{1+k}ug;\\
			\delta \Omega=&2(1+k)g \ .
		\end{align}
		Observe that $m$ and $K$ become $u$-dependent.
		
		\item \textit{$\xi^u=\xi^{r(V)}=0$ and $V^A(z,\zb)\neq0$}
		\begin{align}
			\delta \Omega=&D_AV^A \ .
		\end{align}
		It is zero for rotations but non-zero for general CKV on $S^2$.
		
		\item \textit{$\xi^u=0$, $\xi^{r(V)}=g(z,\bar{z})$ and $V^A\neq0$}
		\begin{align}
			\delta \Phi=&-\frac{2k(2+k)}{1+k}g(z,\bar{z}); && \delta \Psi=-(1+2k)g(z,\bar{z});\\
			\delta m=&\frac{k(k+3)}{(1+k)^2}ug(z,\bar{z}); && \delta K=\frac{2k}{1+k}ug(z,\bar{z});\\
			\delta \Omega=&2(1+k)g(z,\bar{z})+D_AV^A; && \delta \Theta_A=-D_Ag(z,\bar{z});\\
			\delta U_A=&-\frac{k}{1+k}uD_Ag(z,\bar{z}); && \delta N_A=-\frac{k}{(1+k)^2}u^2D_Ag(z,\bar{z}) \ .
		\end{align}
		Apart from $C_{AB}$, all the asymptotic data are generated.
		
	\end{itemize}
	
	
	\subsection{Inhomogeneities}
	\label{sect_inhomog}
	In this section, we consider spacetimes with a central inhomogeneity that manifests itself by non-vanishing subleading terms of $m$ and $K$.
	
	\subsubsection{$m(u,z,\zb)\neq0$}
	In a first step we consider a non-vanishing mass-like term $m(u,z,\zb)\neq0$. Such spacetimes are clearly inhomogeneous and may be anisotropic in case that $m$ is angle dependent.
	
	The asymptotic diffeomorphisms remain exactly as in section \ref{subsec_pureFRW} and the asymptotic data changes only in the following terms
	\begin{align}
		\delta m=&\delta m_{\text{FRW}}+\xi^u\partial_um-(1-2k)m\xi^{r(V)}+V^AD_Am+m\partial_u\xi^u \label{eq:dmgenrotSDyer}\\
		\delta N_A=&\delta N_{A\text{FRW}}+2mD_A\xi^u \label{eq:dNAgenrotpureSDyer} \ .
	\end{align}
	Depending on whether $m$ depends on $u$ and/or $z,\zb$, some terms are present or absent. Furthermore, the orbits of transformations analyzed in section \ref{subsec_pureFRW} do not change because no leading term is affected (by a single transformation) with respect to pure FRW background.
	
	We list only the new terms generated under application of the same transformations as in section \ref{subsec_pureFRW}, and we denote the expressions from that section with the subscribt "FRW".
	
	\begin{itemize}
		\item \textit{$\xi^{r(V)}=V^A=0$ and $\xi^u(u,z,\zb)=f$}
		\begin{align}
			\delta m=&\delta m_{\text{FRW}}+f\partial_um \ .
		\end{align}
		
		\item \textit{$\xi^{r(V)}=V^A=0$ and $\xi^u(u,z,\zb)=f(z,\bar{z})$}
		\begin{align}
			\delta m=&\delta m_{\text{FRW}}+(\partial_um)f(z,\zb); && 
			\delta N_A=\delta N_{A\text{FRW}}+2mD_Af(z,\bar{z}) \ .
		\end{align}

		\item \textit{$\xi^{r(V)}=V^A=0$ and $\xi^u(u,z,\zb)=f(u)$}
		\begin{align}
			\delta m=&\delta m_{\text{FRW}}+(\partial_um)f(u)+m\partial_uf(u) \ .
		\end{align}
		
		\item \textit{$\xi^{r(V)}=0$, $D_AV^A=0$ and $\xi^u(u,z,\zb)=f(u)Y^1_m$}
		\begin{align}
			\delta m=&\xi^u\partial_um+V^AD_Am+m\partial_u\xi^u; && 
			\delta N_A=2mD_A\xi^u \ .
		\end{align}

		\item \textit{$\xi^u=V^A=0$ and $\xi^{r(V)}(z,\zb)=g$}
		\begin{align}
			\delta m=&\delta m_{\text{FRW}}-(1-2k)mg  \ . \label{eq:gnew}
		\end{align}
		
		\item \textit{$\xi^u=\xi^{r(V)}=0$ and $V^A(z,\zb)\neq0$}
		\begin{align}
			\delta m=&\delta m_{\text{FRW}}+V^AD_Am \ .
		\end{align}
		
		\item \textit{$\xi^u=0$, $\xi^{r(V)}=g(z,\bar{z})$ and $V^A\neq0$}
		\begin{align}
			\delta m=&\delta m_{\text{FRW}}-(1-2k)mg(z,\zb)+V^AD_Am \ . \label{eq:gnew2}
		\end{align}
		
	\end{itemize}
	
	Note that the unique different contributions with respect to the same background in the flat case are in (\ref{eq:gnew}) and (\ref{eq:gnew2}). They vanish for $k=\frac12$ (free scalar field) and change sign for $k=1$ (radiation).

	
	\subsubsection{Sultana-Dyer}
	\label{sect_sultana_dyer}
	
	Another seemingly simple example would be the Sultana-Dyer black hole solution \cite{Sultana:2005tp} \footnote{Such solution is formally equivalent to McVittie \cite{McVittie:1933zz} with constant mass (and therefore accretion). It is a conformal Schwarzschild, which corresponds to a black hole embedded in a dust-filled ($k=2$) spatially flat FRW metric \cite{Faraoni:2009uy}.}. However, upon coordinate transformation into Bondi coordinates, we discover that it is not covered by our ansatz, as the scale factor acquires an additional logarithmic $r$ dependence. To leading order in the $\frac{1}{r}$ expansion the Sultana-Dyer solution can be described in our ansatz as a spacetime with non-vanishing constant $m$ and $K$. The Sultana-Dyer metric in the original form \cite{Sultana:2005tp,Faraoni:2009uy} is written as:
	\begin{align}
		\dd s^2=\left(\frac{\eta}{L}\right)^{2k}\left[-\left(1-\frac{2m}{r}\right)\dd\eta^2+\frac{4m}{r}\dd\eta\dd r+\left(1+\frac{2m}{r}\right)\dd r^2+2r^2\gamma_{z\zb}\dd z\dd\zb\right] \ .
	\end{align}
	
	This can be brought into Bondi form. However, there are two ways to write the metric corresponding to two different choices for the radial coordinate:
	\begin{align}
		\dd s^2=&\left(\frac{u+r+2m\log\left(\frac{r}{2m}-1\right)}{L}\right)^{2k}\left[-\left(1-\frac{2m}{r}\right)\dd u^2-2\dd u\dd r+2r^2\gamma_{z\zb}\dd z\dd\zb\right]\label{eq:Sultana_Dyer1}\\
		\dd s^2=&\left(\frac{u+\tilde{r}}{L}\right)^{2k}\left[-\left(1-\frac{2m}{r(\tilde{r})}\right)\dd u^2-2\left(1-\frac{2m}{r(\tilde{r})}\right)\dd u\dd \tilde{r}
		+2r(\tilde{r})^2\gamma_{z\zb}\dd z\dd\zb\right] \ ,\label{eq:Sultana_Dyer2}
	\end{align}
	where $r$ and $\tilde{r}$ are related by:
	\begin{align}
		\tilde{r}=&r+2m\log\left(\frac{r}{2m}-1\right) \ . \label{eq:Sultana_Dyer_r}
	\end{align}
	It is obvious that \eqref{eq:Sultana_Dyer1} is not covered by our ansatz, as the scale factor picks up an additional $r$ dependence. To cover spacetimes that include \eqref{eq:Sultana_Dyer1} we would have to generalize our ansatz for asymptotically FRW spacetimes. This might be necessary to include many interesting examples, but for now we leave this for future investigations. 
	
	In \eqref{eq:Sultana_Dyer2} we have the problem that there is no easy way to expand $r(\tilde{r})$ in orders of $\tilde{r}$ as the inverse of \eqref{eq:Sultana_Dyer_r} is given by the Lambert $W$ function. Assuming that the leading order is proportional to $r$, the Sultana-Dyer solution would correspond, to leading order, to an asymptotically FRW spacetime with constant $m$ and $K=2m$.
	
	The asymptotic diffeomorpisms in this case are given by:
	\begin{align}
		\xi^{A(1)}=&\xi^{A(1)}_{FRW};\qquad \xi^{A(2)}=mD^A\xi^u;\qquad \xi^{r(0)}=\xi^{r(0)}_{FRW}\\
		\xi^{r(1)}=&\xi^{r(1)}_{FRW}-\frac{m}{2(1+k)}D_AD^A\xi^u \ .
	\end{align}
	The asymptotic data changes in the following terms. Again we only list the terms that transform differently to the pure FRW case:
	\begin{align}
		\delta m=&\delta m_{FRW}+m\partial_u\xi^u+2m\partial_u\xi^{r(0)}-(1-2k)m\xi^{r(V)}\label{eq:dm_Sultana_Dyer}\\
		\delta K=&\delta K_{FRW}+2m\partial_u\xi^u+4km\xi^{r(V)}\label{eq:dK_Sultana_Dyer}\\
		\delta N_A=&\delta N_{A\ FRW}+\frac{m}{1+k}D_AD_BD^B\xi^u+2mD_A\xi^u+2mD_A\xi^{r(0)} \ .
	\end{align}
	By comparing \eqref{eq:dm_Sultana_Dyer} and \eqref{eq:dK_Sultana_Dyer} we see that after a general transformation, the new metric does not look like a Sultana-Dyer spacetime, as $K=2m$ is no longer true. 
	
	
	\subsection{Cosmological perturbations}
	\label{sect_cosmopert}
	
	Cosmological perturbations are usually classified into scalar, vector and tensor modes according to their transformation behaviour under spatial rotations. It turns out that Bondi coordinates and Bondi gauge are suited to easily identify the cosmological modes.\footnote{For a more detailed treatment of the cosmological perturbations and their physical relevance we recommend \cite{Mukhanov:2005sc}.} The perturbed spatially flat FRW metrics we consider \footnote{Note that the more general class of metrics would contain $r^2(\gamma_{AB}+q_{AB})$. This would allow for leading order tensor degrees of freedom, but it is incompatible with the restriction of $V^A$ to be a CKV and the Bondi gauge.} can be written in Bondi coordinates as
	\begin{align}
		\dd s^2=&a^{2}\left\{-\left(1-\Phi\right)\dd u^2-2\left(1-\Psi\right)\dd u\dd r-2r\Theta_A\dd u\dd x^A\right. \nonumber \\
		&\left.+\left((1+\Omega)r^2\gamma_{AB}+rC_{AB}\right)\dd x^A\dd x^B\right\} \ ,
		\label{eq:fullpertspatflatFLRW}
	\end{align}
	where all the coefficients are small (perturbations), have a priori arbitrary dependence on $u,z,\zb$ and their dependence on $r$ is only restricted to be at most $\mathcal{O}(1)$ at large $r$.
	
	Written in this form, $\Phi$, $\Psi$, $\Omega$ and $C^A_A$  transform as scalars, $\Theta_A$ as a vector and the traceless part of $C_{AB}$ as a tensor under spatial rotations. This can easily be checked by using the definition of spatial rotations in Bondi coordinates that are given in \eqref{eq:GKVrotations} and the general transformation laws \eqref{eq:dNgenrot}-\eqref{eq:dNAgenrot}. Note that we can still create additional scalar, vector and tensor modes by adding or contracting with a covariant derivative.\footnote{For example, for a scalar $E$ the term $D_A E$ would transform like a vector, while for a tensor $J_{AB}$ the divergence $D^AJ_{AB}$ transforms like a vector as well.} This is due to the fact that we are still using an off-shell formalism. Once we enforce the equations of motion and make restrictions for the matter content we allow in the energy-momentum tensor, the degrees of freedom should be completely fixed to two scalar modes, two vector modes and two tensor modes. But even if we assume that there are no hidden additional modes in \eqref{eq:fullpertspatflatFLRW}, there are still four scalar modes instead of two. This is due to the fact that we have not imposed the strong Bondi gauge yet which fixes the gauge completely and restricts $\Omega=0$ or $\Omega$ being non-dynamical and $C^A_A=0$.
	
	In summary we can say that the modes in our ansatz \eqref{eq:fullpertspatflatFLRW} represent the maximal number of degrees of freedom we can have for each perturbation. Nevertheless, only an on-shell treatment using the energy momentum tensor determined by a theory (for example General Relativity) can tell us whether those modes are actually real. Such on-shell treatment was performed in Cartesian coordinates long ago \cite{Mukhanov:2005sc} but not in Bondi coordinates in a compatible way with the $r$-expansion. It is extremely interesting because then one would be able to determine how exactly on-shell asymptotic diffeormorphisms change the cosmological modes for different observers, but also challenging and we leave it for future studies.
	
	We will now work out three particularly simple cases off-shell and leave their on-shell treatment, as well as more complicated examples for future treatment.


	\subsubsection{Scalar mode background}
	
	Let us take a background with only one leading order scalar mode $\Phi\neq0$ and the rest of the asymptotic data to vanish. In such a case, the asymptotic diffeomorphisms remain exactly as in section \ref{subsec_pureFRW} and the asymptotic data reads as follows 
	\begin{align}
		\delta\Phi=&\delta \Phi_{FRW}+V^AD_A\Phi+\xi^u\partial_u\Phi+2k\Phi\xi^{r(V)}+2\Phi\partial_u\xi^u\label{eq:dNgenrotsmode}\\
		\delta m=&\delta m_{FRW}+k\Phi\xi^u-ku\Phi\xi^{r(V)}+k\Phi\xi^{r(0)}+\frac12\xi^{A(1)}D_A\Phi\label{eq:dmgenrotsmode}\\
		\delta U_A=&\delta U_{A \ FRW}+\Phi D_A\xi^u \ . \label{eq:dUAgenrotsmode}
	\end{align}
	Note that unless $\xi^u$ is angle dependent, the background scalar mode does not generate any mixed modes by itself apart from the pure FRW background. In such a case, still the vector-like contribution could be revealed to be "fake" by the equations of motion in an on-shell treatment. 
	
	Let us note that, for a constant supertranslation $\xi^u=f$, the only non vanishing components for pure FRW background are $\delta m_{FRW}$ and $\delta K_{FRW}$. Moreover, both were just constant and therefore can be regarded just as a background shift. Nevertheless, the $\Phi$-background mode evolves at leading order in $r$ picking up a linear $f\partial_u\Phi$ contribution and generates to subleading order a $u$-dependent contribution $m(u)-\delta m_{FRW}=k\Phi f+k\Phi\xi^{r(0)}$ in the supertranslated frame.
	
	
	\subsubsection{Vector mode background}
	
	Next, we analyze a background with two leading order vector modes $\Theta_A\neq0$ and the rest of the asymptotic data to vanish. In such a case, the asymptotic diffeomorphisms are modified
	\begin{align}
		\xi^{A(1)}=&\xi^{A(1)}_{FRW};\qquad
		\xi^{A(2)}=\xi^{A(2)}_{FRW}=0\\
		\xi^{r(0)}=&\xi^{r(0)}_{FRW}-\frac{1}{2(1+k)}\Theta^AD_A\xi^u=\xi^{r(0)}_{FRW}+\tilde{\xi}^{r(0)}  \\
		\xi^{r(1)}=&\xi^{r(1)}_{FRW}-\frac{ku}{2(1+k)^2}\Theta^AD_A\xi^u=\xi^{r(1)}_{FRW}+\tilde{\xi}^{r(1)} \ ,
	\end{align}
	and the asymptotic data transforms as follows
	\begin{align}
		\delta\Phi=&\delta \Phi_{FRW}-2\partial_u\tilde{\xi}^{r(0)}+2\Theta_A\partial_u\xi^{A(1)}\label{eq:dNgenrotvmode}\\
		\delta m=&\delta m_{FRW}-k\tilde{\xi}^{r(0)}-\partial_u\tilde{\xi}^{r(1)} \label{eq:dmgenrotvmode} \\
		\delta K=&\delta K_{FRW}-\Theta_A\xi^{A(1)}-2k\tilde{\xi}^{r(0)}\label{eq:dFgenrotvmode}\\
		\delta C_{AB}=&\delta C_{AB \ FRW}+2\gamma_{AB}(1+k)\tilde{\xi}^{r(0)}+\Theta_AD_B\xi^u+\Theta_BD_A\xi^u \label{eq:dCABgenrotvmode} \\
		\delta \Theta_A=&\delta \Theta_{A \ FRW}+ V^BD_B\Theta_A+\xi^u\partial_u\Theta_A+(1+2k)\Theta_A\xi^{r(V)} \nonumber \\
		&+\Theta_BD_AV^B+\Theta_A\partial_u\xi^u\label{eq:dAAgenrotvmode}\\
		\delta U_A=&\delta U_{A \ FRW}+2k\Theta_A\xi^u+(1+2k)\Theta_A\xi^{r(0)}-2k\xi^{r(V)}u\Theta_A\nonumber\\
		&+\xi^{B(1)}D_B\Theta_A+\Theta_BD_A\xi^{B(1)}-D_A\tilde{\xi}^{r(0)}\label{eq:dUAgenrotvmode}\\
		\delta N_A=&\delta N_{A \ FRW}
		-D_A\tilde{\xi}^{r(1)}+2k\Theta_A\left(u^2\xi^{r(V)}-u(\xi^{r(0)}+\xi^u)+\xi^{r(1)}\right) \nonumber \\
		&+\Theta_A\xi^{r(1)} \ . \label{eq:dNAgenrotvmode}
	\end{align}
	
	It it easy to realize that, unless we apply an angle dependent supertranslation, i.e. $\xi^u=f(z,\zb)$, there is no mode mixing apart from the FRW background shift and the effect of the background vector modes $\Theta_A$ is restricted to the vector components. That is because in such a case $D_A\xi^u=0$ leads also to $\xi^{A(1)}=\tilde{\xi}^{r(0)}=\tilde{\xi}^{r(1)}=0$.
	
	
	\subsubsection{Tensor mode background}
	
	Finally, we choose a background with two subleading order gravitational modes $C_{AB}\neq0$ ($C^A_A=0$, $C_{zz},C_{\bar{z}\bar{z}}\neq0$) and the rest of the asymptotic coefficients to vanish. In such a case, the asymptotic diffeomorphisms are modified
	\begin{align}
		\xi^{A(1)}=&\xi^{A(1)}_{FRW};\quad 
		\xi^{A(2)}=\frac12 C^{BA}D_B\xi^u;\quad
		\xi^{r(0)}=\xi^{r(0)}_{FRW}  \\
		\xi^{r(1)}=&\xi^{r(1)}_{FRW}-\frac{1}{4(1+k)}D_{A}[C^{BA}D_B\xi^u]=\xi^{r(1)}_{FRW}+\tilde{\xi}^{r(1)} \ ,
	\end{align}
	and the asymptotic data reads as follows
	\begin{align}
		\delta m=&\delta m_{FRW}-\partial_u\tilde{\xi}^{r(1)} \label{eq:dmgenrotgmode} \\
		\delta C_{AB}=&\delta C_{AB \ FRW}+\xi^u\partial_uC_{AB}+V^CD_C C_{AB}+C_{AC}D_BV^C\nonumber\\
		&+C_{BC}D_AV^C+(1+2k)C_{AB}\xi^{r(V)} \label{eq:dCABgenrotgmode} \\
		\delta U_A=&\delta U_{A \ FRW}+C_{AB}\partial_u\xi^{B(1)}+\partial_u\xi^{B(2)}\label{eq:dUAgenrotgmode}\\
		\delta N_A=&\delta N_{A \ FRW}
		-D_A\tilde{\xi}^{r(1)}+C_{AB}\partial_u\xi^{B(2)} \ . \label{eq:dNAgenrotgmode}
	\end{align}
	
	The same comments as in the end of the previous subsection apply here, meaning that unless we apply an angle dependent supertranslation there is no mode mixing apart from the FRW background shift and the effect of the background gravitational modes $C_{zz}$ and $C_{\zb\zb}$ is restricted to themselves. That is because in such a case $D_A\xi^u=0$ leads also to $\xi^{A(1)}=\xi^{A(2)}=\tilde{\xi}^{r(1)}=0$. 
	

	\section{Summary, conclusions and future research}
	\label{sect_conc}
	
	In this paper we have studied asymptotically decelerating spatially flat FRW spacetimes at future null infinity $\mathcal{I}^+$ in a gauge similar to Bondi. More concretely, we have performed an off-shell analysis where the Einstein equations have been only used indirectly, in order to further motivate the ansatz (\ref{eq:asymptpertspatflatFLRW}). Following the standard procedure in asymptotically flat spacetimes, we have obtained asymptotic diffeomorphisms consistent with the global symmetries of FRW, and we have investigated their effect on diverse backgrounds. Our main goals were to obtain the more general transformations mapping asymptotically spatially flat FRW spacetimes among themselves and to investigate how such diffeomorphisms transform the physically relevant asymptotic metric coefficients. Both fit within the aim of addressing the still missing \textit{asymptotic symmetry corner of the cosmological infrared triangle} \cite{Strominger:2017zoo}.

	
	\subsubsection*{Summary of results}
	
	Let us summarize our main findings:
	
	\begin{itemize}
		\item We have motivated and defined the class of metrics to be considered asymptotically decelerating spatially flat FRW at future null infinity $\mathcal{I}^+$ (\ref{eq:asymptpertspatflatFLRW}). Next, we have obtained the supertranslation-like diffeomorphisms that act consistently on them (\ref{eq:Stranslac}), and then further extend to include superrotation-like transformations (\ref{eq:Srotacc}) together with their effect on the asymptotic data \eqref{eq:dNgenrot}-\eqref{eq:dNAgenrot}. Particularly, we have checked that these transformations are consistent with the GKV of FRW and recovered the asymptotically flat case in the limit $k=0$. Along the way, we have adopted the normal Bondi gauge, such that the gauge was not totally fixed, in order to allow for more general metrics. Nevertheless, we have explained in detail how to fix the remaining degree of freedom by means of the so-called strong Bondi gauge and why it is important for a future on-shell treatment of cosmological perturbations.

		\item We have investigated how these tranformations act on a pure FRW background and noticed that, contrary to the flat case, not only $C_{zz}$ and $C_{\zb\zb}$ but all the asymptotic data can be generated out of the vacuum. This partially reflects the still off-shell treatment we developed but also the fact that vector and scalar modes are allowed to be present in a cosmological background. Besides, we have also studied the compatibility between sub-classes of transformations which do not create leading order asymptotic coefficients.
		
		\item We have computed the extra contributions due to the presence of non-vanishing $m$ in the background. Furthermore, we have realized that central inhomogeneities are more involved to describe than Schwarzschild in the flat case. Its conformal counterpart, the Sultana-Dyer black hole background, has been explored and we have realized that it is not covered by our ansatz (\ref{eq:asymptpertspatflatFLRW}), although for large $r$ could be simulated by  $2m=K=\text{constant}$.
		
		\item Finally, we have considered some particularly simple cosmologically perturbed backgrounds and analyzed how the perturbations affect the remaining asymptotic data under the action of the asymptotic diffeomorphisms. We have admitted that a full analysis would require an on-shell treatment but we have given hints on how it could be performed, noticed subtleties and necessary details to be considered.
	\end{itemize}

	
	\subsubsection*{Open questions and future directions}
	
	We would like to conclude by listing some especially relevant future research guidelines.
	
	\begin{itemize}
		\item In order to make our analysis useful for the precise description of physical data, we have to dive into an on-shell description. Although we used Bondi gauge and Einstein equations to motivate the asymptotic metrics (\ref{eq:asymptpertspatflatFLRW}), there are more coefficients in the off-shell $r$-expansion  (\ref{eq:asymptpertspatflatFLRW}) than the actual on-shell degrees of freedom corresponding to two scalar modes, two vector modes and two gravitational modes (\ref{eq:fullpertspatflatFLRW}). The implementation of the equations of motion should relate several of these coefficients, providing with the right number of degrees of freedom. Establishing a classification of asymptotic metrics based on more strict falloffs for the allowed matter content (energy momentum tensor) seems to be the first reasonable step. Furthermore, exploring the compatibility of our formalism with the more geometric one of \cite{Bonga:2020fhx} might clarify the path to follow in order to find the on-shell description.
		
		\item Even though exact Sultana-Dyer is not a physical solution \cite{Faraoni:2009uy,Faraoni:2013aba}, its inclusion would require to modify our ansatz (\ref{eq:asymptpertspatflatFLRW}) because it is not described by a $\frac1r$ expansion. It might be possible to obtain inspiration from polyhomogeneous expansions in flat spacetime \cite{Godazgar:2020peu}. In general, it would be interesting to broaden the analysis of section \ref{sect_inhomog} to more involved cosmological black hole solutions \cite{Faraoni:2013aba}, as well as to the asymptotically spatially flat FRW LTB (Lemaitre-Tolman-Bondi) metrics formulated in \cite{Firouzjaee:2008gs}. The latter are better suited to describe the collapse of a spherical mass distribution with over density within a cosmological setting than Sultana-Dyer and McVittie's black hole. Furthermore, one could also try to apply our analysis to flat Kerr-Newmann black holes ($k=0$) \cite{PhysRevD.73.084023,article} and Kerr-Newmann embedded in cosmology \cite{Vaidya:1977zza}.

		\item A beneficial feature of our off-shell analysis is that it is not restricted to Einstein gravity. It allows for an on-shell implementation in general gravity theories which have decelerated spatially flat FRW as solutions.

		\item In order to explore our formalism in spacetimes with leading order anisotropies \footnote{Note that, due to $\delta g_{rA}=0$, the only manner to introduce anisotropies is through generation of diagonal terms $r^2q_{AB}$ deforming the off-diagonal round metric.} (e.g. Bianchi Type I) and leading order diagonal perturbations to the round sphere metric, we have to allow for $V^A\in \text{Vect}(S^2)$ instead of restricting $V^A$ to be a CKV on $S^2$. The gauge we adopted is not suited for this task, requiring a complete reformulation of the analysis presented here.

		\item The asymptotic diffeomorphisms we obtained (\ref{eq:Srotacc}) are considerably more complicated than in the flat case and depend explicitly on several metric coefficients. We did not identify their algebra neither for concrete special orbits nor in full generality. However, it is essential to understand the structure they describe in order to make the connection with black hole entropy, fluid-gravity duality and membrane models embedded in a cosmological setting. Three main features make it more complicated than in asymptotically flat spacetimes: the asymptotic diffeomorphisms depend on several asymptotic data, $\xi^{u}$ is $u$-dependent and $\xi^{r(V)}$ is another parameter to deal with in case we do not implement strong Bondi gauge.
		
		\item Our analysis and diffeomorphisms are infinitesimal. Ultimately, it would be desirable to extend them to be finite.
	\end{itemize}
	
	
	\vskip1em
	\subsubsection*{Acknowledgments}
	
	We would like to thank I.~Sachs for enlightening discussions and valuable suggestions regarding this work and general support. TH thanks the Hans-B\"ockler-Stiftung of the German Trade Union Confederation (DGB) for financial support. 
	MER is grateful to M.~Sheikh-Jabbari for early stage explanations, as well as to I.~Kharag for proofreading this paper. The work of MER was supported by the DFG Excellence Cluster ORIGINS.
	
	
	\appendix
	
	
	\section{Einstein tensor}
	\label{einsteintensor}
	In this part of the appendix we write down the Einstein tensor in the $\frac{1}{r}$ expansion. The Einstein tensor can be separated in the following way:
	\begin{align}
		G_{\mu\nu}=&\bar{G}_{\mu\nu}+\delta G_{\mu\nu}\label{eq:Gmn} \ ,
	\end{align}
	where $\bar{G}_{\mu\nu}$ is the Einstein tensor of the pure FRW background whose non-vanishing components are given by:
	\begin{align}
		\bar{G}_{uu}=&\frac{3 k^2}{r^2}+\Op(r^{-3})\\
		\bar{G}_{ur}=&\frac{3 k^2}{r^2}+\Op(r^{-3})\\
		\bar{G}_{rr}=&\frac{2 k (k+1)}{r^2}+\Op(r^{-3})\\
		\bar{G}_{z\zb}=&-\gamma_{z\zb}k(k-2)+\Op(r^{-1}) \ .
	\end{align}
	$\delta G_{\mu\nu}$ are the terms that are generated by the perturbations in \eqref{eq:asymptpertspatflatFLRW}. We expand this in $\frac{1}{r}$ in the following way:
	\begin{align}
		\delta G_{\mu\nu}=&r\delta G^{(-1)}_{\mu\nu}+\delta G^{(0)}_{\mu\nu}+\frac{1}{r}\delta G^{(1)}_{\mu\nu}+\frac{1}{r^2}\delta G^{(2)}_{\mu\nu}+\Op(r^{-3}) \ .
	\end{align}
	As the $\delta G^{(i)}_{\mu\nu}$ s are in general very lengthy and complicated, we write down only the terms that are of first order in the perturbations. Therefore, they are only valid in the case where these perturbations are very small. To analyse the $\frac{1}{r}$ behaviour of the Einstein tensor, however, this is sufficient:
	\begin{align}
		\delta G^{(0)}_{uu}=&-\partial^2_u\Omega\\
		\delta G^{(1)}_{uu}=&-(1+k)\partial_u\Phi+2\partial_u\Psi+(3+2k)\partial_u\Omega+\partial_uD_A\Theta^A\\
		\delta G^{(2)}_{uu}=&(1+4k)\Phi-2(1+2k)\Psi-\Omega-2(1+k)D_A\Theta^A-D_AD^A\Phi\nonumber\\
		&+2D_AD^A\Psi-D_AD^A\Omega+2\partial_uK-2(1+k)\partial_u m+ku\partial_u\Phi\nonumber\\
		&-2ku\partial_u\Omega+\partial_uD_AU^A-\partial_uh^A_A\\
		\delta G^{(1)}_{ur}=&2(1+k)\partial_u\Omega\\
		\delta G^{(2)}_{ur}=&(1+k)^2\Phi-(2+2k+k^2)\Psi-\Omega-\frac32D_A\Theta^A-2kD_A\Theta^A\nonumber\\
		&+D_AD^A\Psi-D_AD^A\Omega-2ku\partial_u\Omega\\
		\delta G^{(1)}_{rA}=&(1+k)(\Theta_A-D_A\Psi)\\
		\delta G^{(2)}_{rA}=&(1+2k)U_A+ku(D_A\Psi-\Theta_A)-\frac12D^BC_{AB}-\frac12(3+2k)D_A K\\
		\delta G^{(0)}_{uA}=&\frac12\left(\partial_uD_A\Psi-\partial_uD_A\Omega-\partial_u\Theta_A\right)\\
		\delta G^{(1)}_{uA}=&(2-k)k\Theta_A-kD_A\Phi+\frac12D_AD_B\Theta^B-\frac12D_BD^B\Theta_A-\partial_uU_A\nonumber\\
		&+\frac12\partial_uD^BC_{AB}+\frac{1}{2}\partial_uD_AK\\
		\delta G^{(0)}_{zz}=&-(1+2k)D_z\Theta_z+D_zD_z\Psi+k\partial_uC_{zz}\\
		\delta G^{(1)}_{zz}=&(2-k)kC_{zz}-2kD_zU_z+2kuD_z\Theta_z+D_zD_zK-ku\partial_uC_{zz}\nonumber\\
		&-(1-k)\partial_u h_{zz}\\
		\delta G^{(-1)}_{z\zb}=&-(1+k)\gamma_{z\zb}\partial_u\Omega\\
		\delta G^{(0)}_{z\zb}=&\gamma_{z\zb}\left((2-k)k\Omega+2k\Psi-k^2\Phi-\partial_uK+ku\partial_u\Omega\right)-2D_zD_{\zb}\Psi\nonumber\\
		&+(1+2k)\left(D_{\zb}\Theta_z+D_z\Theta_{\zb}\right)\\
		\delta G^{(1)}_{z\zb}=&\gamma_{z\zb}\left(2k(2-k)m-K-2ku((1-k)\Phi+\Psi+2(2-k)\Omega+u\partial_u\Omega)\right)\nonumber\\
		&+k\left(D_{\zb}U_z+D_zU_{\zb}\right)-ku\left(D_{\zb}\Theta_z+D_z\Theta_{\zb}\right)-D_zD_{\zb}K\nonumber\\
		&+(1-k)\partial_uh_{z\zb} \ .
	\end{align}
	
	
	\section{Lie derivatives}
	\label{lieder}
	
	In this appendix we collect the Lie derivatives resulting from the action of supertranslations (\ref{eq:astrans}) and superrotations (\ref{eq:superrot_ansatz}) in the asymptotically spatially flat FRW metric (\ref{eq:asymptpertspatflatFLRW}). Both cases share the Lie derivative
	\begin{align}
		a^{-2}\mathcal{L}_{\xi}g_{rr}=&-2\left(1-\Psi-\frac{K}{r}+\op{O}(r^{-2})\right)\partial_r\xi^u \ ,
		\label{eq:bondigauge1}
	\end{align}
	which translates into $\partial_r\xi^u=0$ to verify the Bondi gauge. Therefore, we consider only $\xi^u(u,z,\zb)$.
	
	
	\subsection{Supertranslations}
	\label{liederstrans}
	
	\begin{align}
		a^{-2}\Lie{\xi}{g_{uu}}=&+\left[\xi^u\partial_u\Phi-2(1-\Psi)\partial_u\xi^{r(0)}-2(1-\Phi)\partial_u\xi^u+2\Theta_A\partial_u\xi^{A(1)}\right] \nonumber \\
		&+\frac{2}{r}\left[\xi^u\partial_u m-k(1-\Phi)\xi^u-k(1-\Phi)\xi^{r(0)}+\frac12\xi^{A(1)}D_A\Phi\right. \nonumber\\
		&+K\partial_u\xi^{r(0)}-(1-\Psi)\partial_u\xi^{r(1)}+m\partial_u\xi^u+U_A\partial_u\xi^{A(1)}\nonumber \\
		&\left.+\Theta_A\partial_u\xi^{A(2)}\right]+\Op(r^{-2})\label{eq:Lieuu}\\
		a^{-2}\Lie{\xi}{g_{ur}}=&\left[\xi^u\partial_u\Psi+(\Psi-1)\partial_u\xi^u\right]+\frac{1}{r}\left[\xi^u\partial_u K+K\partial_u\xi^u+\xi^{A(1)}D_A\Psi\right. \nonumber \\
		&\left.-\Theta_A\xi^{A(1)}-2k(1-\Psi)\left(\xi^u+\xi^{r(0)}\right)\right]+\Op(r^{-2})\label{eq:Lieur}\\
		a^{-2}\Lie{\xi}{g_{uA}}=&r\left[\xi^u\partial_u\Theta_A+\Theta_A\partial_u\xi^u+(1+\Omega)\partial_u\xi^{(1)}_A\right]+\left[(2k\Theta_A+\partial_u U_A)\xi^u \right. \nonumber \\
		&\left.+(1+2k)\Theta_A\xi^{r(0)}+\xi^{B(1)}D_B\Theta_A+\Theta_BD_A\xi^{B(1)}-(1-\Psi)D_A\xi^{r(0)}\right. \nonumber\\
		&\left.-(1-\Phi)D_A\xi^u+U_A\partial_u\xi^u+C_{AB}\partial_u\xi^{B(1)}+(1+\Omega)\partial_u\xi_A^{(2)}\right]\nonumber \\
		&+\frac{1}{r}\left[\xi^u\partial_u N_A+N_A\partial_u\xi^u+\xi^{B(1)}D_BU_A+U_BD_A\xi^{B(1)}+\xi^{B(2)}D_B\Theta_A\right. \nonumber \\
		&+\Theta_BD_A\xi^{B(2)}+KD_A\xi^{r(0)}-(1-\Psi)D_A\xi^{r(1)}+2mD_A\xi^u \nonumber\\
		&+2kU_A(\xi^{r(0)}+\xi^u)+2k\Theta_A\left(-u(\xi^{r(0)}+\xi^u)+\xi^{r(1)}\right) \nonumber \\
		&\left.+\Theta_A\xi^{r(1)}+C_{AB}\partial_u\xi^{B(2)}+h_{AB}\partial_u\xi^{B(1)}\right]+\Op(r^{-2})\label{eq:LieuA}\\
		a^{-2}\Lie{\xi}{g_{rA}}=&-\gamma_{AB}\left((1+\Omega)\xi^{B(1)}+(1-\Psi)D^B\xi^u\right)\nonumber\\
		&+\frac1r\left(KD_A\xi^u-2(1+\Omega)\xi_A^{(2)}-C_{AB}\xi^{B(1)}\right)+\Op(r^{-2}) 
		\label{eq:liestransrA}\\
		a^{-2}\Lie{\xi}{g_{AB}}=&r^2F_{AB}+rS_{AB}+K_{AB} \label{eq:LieAB1}
	\end{align}
	with 
	\begin{align}
		F_{AB}=&\gamma_{AB}\xi^u\partial_u\Omega \ , \nonumber \\
		S_{AB}=&2(1+\Omega)\gamma_{AB}((1+k)\xi^{r(0)}+k\xi^u)+\gamma_{AB}\xi^{C(1)}D_C \Omega+\Theta_AD_B\xi^u\nonumber\\
		&+\Theta_B D_A\xi^u+(1+\Omega)(D_A\xi_B^{(1)}+D_B\xi_A^{(1)})+\xi^u\partial_u C_{AB} \ , \nonumber \\
		K_{AB}=&-2k(1+\Omega)\gamma_{AB}\left(u\xi^{r(0)}+u\xi^u\right)+2(1+k)(1+\Omega)\gamma_{AB}\xi^{r(1)}\nonumber\\
		&+\gamma_{AB}\xi^{C(2)}D_C\Omega+\xi^u\partial_u h_{AB}+U_AD_B\xi^u+U_BD_A\xi^u+C_{AC}D_B\xi^{C(1)}\nonumber\\
		&+C_{BC}D_A\xi^{C(1)}+\xi^{C(1)}D_C C_{AB}+(1+\Omega)(D_A\xi_B^{(2)}+D_B\xi_A^{(2)}) . \label{eq:LieAB11}
	\end{align}
	
	
	\subsection{Superrotations}
	\label{liedersrot}
	
	\begin{align}
		a^{-2}\Lie{\xi}{g_{uu}}=&2r\left[\Theta^A\partial_u V_A-(1-\Psi)\partial_u\xi^{r(V)}\right] \nonumber\\
		&+\left[V^AD_A\Phi+\xi^u\partial_u\Phi+2U_A\partial_uV^A-2(1-\Psi)\partial_u\xi^{r(0)}-2k(1-\Phi)\xi^{r(V)}\right. \nonumber\\
		&\left.+2K\partial_u\xi^{r(V)}-2(1-\Phi)\partial_u\xi^u+2\Theta_A\partial_u\xi^{A(1)}\right] \nonumber \\
		&+\frac{2}{r}\left[\xi^u\partial_u m-k(1-\Phi)\xi^u-\left((1-2k)m-ku(1-\Phi)\right)\xi^{r(V)}\right. \nonumber\\
		&-k(1-\Phi)\xi^{r(0)}+V^AD_Am+\frac12\xi^{A(1)}D_A\Phi+K\partial_u\xi^{r(0)}-(1-\Psi)\partial_u\xi^{r(1)}\nonumber \\
		&\left.+m\partial_u\xi^u+U_A\partial_u\xi^{A(1)}+\Theta_A\partial_u\xi^{A(2)}+N_A\partial_uV^A\right]+\Op(r^{-2}) \label{eq:uusrot}\\
		a^{-2}\Lie{\xi}{g_{ur}}=&\left[(1+2k)(\Psi-1)\xi^{r(V)}+V^A\partial_A\Psi+\xi^u\partial_u\Psi+(\Psi-1)\partial_u\xi^u\right] \nonumber \\
		&+\frac{1}{r}\left[\xi^u\partial_u K+V^AD_AK+K\partial_u\xi^u+\xi^{A(1)}D_A\Psi-\Theta_A\xi^{A(1)}\right. \nonumber \\
		&\left.+2k(1-\Psi)\left(u\xi^{r(V)}-\xi^u-\xi^{r(0)}\right)+2kK\xi^{r(V)}\right]+\Op(r^{-2})\label{eq:ursrot} \\
		a^{-2}\Lie{\xi}{g_{uA}}=&(1+\Omega)\partial_uV_A r^2+r\left[(1+2k)\Theta_A\xi^{r(V)}+V^B\partial_B \Theta_A+\Theta_B\partial_AV^B\right. \nonumber\\
		&\left.-(1-\Psi)\partial_A\xi^{r(V)}+C_{AB}\partial_uV^B+\xi^u\partial_u\Theta_A+\Theta_A\partial_u\xi^u+(1+\Omega)\partial_u\xi^{(1)}_A\right] \nonumber \\
		&+\left[(2k\Theta_A+\partial_u U_A)\xi^u+(1+2k)\Theta_A\xi^{r(0)}+2k\xi^{r(V)}(U_A-u\Theta_A)\right. \nonumber\\
		&+V^BD_BU_A+\xi^{B(1)}D_B\Theta_A+\Theta_BD_A\xi^{B(1)}+U_BD_AV^B \nonumber\\
		&-(1-\Psi)D_A\xi^{r(0)}+KD_A\xi^{r(V)}-(1-\Phi)D_A\xi^u+h_{AB}\partial_uV^B \nonumber \\
		&\left.+U_A\partial_u\xi^u+C_{AB}\partial_u\xi^{B(1)}+(1+\Omega)\partial_u\xi_A^{(2)}\right]\nonumber \\
		&+\frac{1}{r}\left[\xi^u\partial_u N_A+N_A\partial_u\xi^u+V^BD_BN_A+N_BD_AV^B-(1-2k)N_A\xi^{r(V)}\right. \nonumber \\
		&+\xi^{B(1)}D_BU_A+U_BD_A\xi^{B(1)}+\xi^{B(2)}D_B\Theta_A+\Theta_BD_A\xi^{B(2)}+KD_A\xi^{r(0)} \nonumber\\
		&-(1-\Psi)D_A\xi^{r(1)}+2mD_A\xi^u+2kU_A(\xi^{r(0)}+\xi^u-u\xi^{r(V)}) \nonumber \\
		&+2k\Theta_A\left(u^2\xi^{r(V)}-u(\xi^{r(0)}+\xi^u)+\xi^{r(1)}\right)+\Theta_A\xi^{r(1)}+C_{AB}\partial_u\xi^{B(2)} \nonumber\\
		&\left.+h_{AB}\partial_u\xi^{B(1)}\right]+\Op(r^{-2}) \label{eq:uAsrot} \\
		a^{-2}\Lie{\xi}{g_{rA}}=&-\gamma_{AB}\left((1+\Omega)\xi^{B(1)}+(1-\Psi)D^B\xi^u\right)\nonumber\\
		&+\frac1r\left(KD_A\xi^u-2(1+\Omega)\xi_A^{(2)}-C_{AB}\xi^{B(1)}\right)+\Op(r^{-2})
		\label{eq:liesrotrA}\\
		a^{-2}\Lie{\xi}{g_{AB}}=&r^2F_{AB}+rS_{AB}+K_{AB} \label{eq:LieAB2}
	\end{align}
	with 
	\begin{align}
		F_{AB}=&\gamma_{AB}(V^CD_C\Omega+2(1+k)\xi^{r(V)}+\xi^u\partial_u\Omega)+(1+\Omega)(D_AV_B+D_BV_A)\nonumber \ , \\
		S_{AB}=&2(1+\Omega)\gamma_{AB}((1+k)\xi^{r(0)}-ku\xi^{r(V)}+k\xi^u)+\gamma_{AB}\xi^{C(1)}D_C \Omega\nonumber\\
		&+(1+\Omega)(D_A\xi_B^{(1)}+D_B\xi_A^{(1)})+\Theta_AD_B\xi^u+\Theta_B D_A\xi^u+(1+2k)C_{AB}\xi^{r(V)}\nonumber\\
		&+C_{AC}D_BV^C+C_{BC}D_AV^C+V^CD_C C_{AB}+\xi^u\partial_u C_{AB}\nonumber \ , \\
		K_{AB}=&2k(1+\Omega)\gamma_{AB}\left(u^2\xi^{r(V)}-u\xi^{r(0)}-u\xi^u\right)+2(1+k)(1+\Omega)\gamma_{AB}\xi^{r(1)}\nonumber\\
		&+\gamma_{AB}\xi^{C(2)}D_C\Omega+\xi^u\partial_u h_{AB}+h_{AC}D_BV^C+h_{BC}D_AV^C+V^CD_Ch_{AB}\nonumber\\
		&+2kh_{AB}\xi^{r(V)}+U_AD_B\xi^u+U_BD_A\xi^u+C_{AC}D_B\xi^{C(1)}+C_{BC}D_A\xi^{C(1)}\nonumber\\
		&+\xi^{C(1)}D_C C_{AB}+(1+\Omega)(D_A\xi_B^{(2)}+D_B\xi_A^{(2)})\label{eq:LieAB11srot} \ .
	\end{align}
	
	
	\clearpage
	\nocite{*}
	\bibliography{references}
	\bibliographystyle{JHEP}
	
	
\end{document}